\documentclass[trackchanges,twocolumn]{aastex62}
\usepackage{natbib}
\usepackage{graphicx}
\usepackage{amsmath}

\newcommand{\citeth}[1]{(\citeauthor{#1}\ \citeyear{#1})}
\newcommand{\citethnop}[1]{\citeauthor{#1}\ \citeyear{#1}}

\def \mkms {{\rm \; km\;s^{-1}}}
\def \msunperyr {{\; M_{\odot}\rm \;yr^{-1}}}

\begin{document}
\title{A VLT/FORS2 Narrowband Imaging Search for \ion{Mg}{2} Emission Around $\lowercase{z}\sim0.7$ Galaxies }
\author{Ryan Rickards Vaught}
\affiliation{Department of Physics, University of California, San Diego, 9500 Gilman Dr., La Jolla, CA 92093, USA}
\affiliation{Department of Astronomy, San Diego State University, San Diego, CA 92182, USA}

 \author{Kate H. R. Rubin}
 \affiliation{Department of Astronomy, San Diego State University, San Diego, CA 92182, USA}
 
 \author{Fabrizio Arrigoni Battaia }
 \affiliation{European Southern Observatory, Karl-Schwarzschild-Str. 2, D-85748 Garching bei M\"unchen, Germany}

 \author{J. Xavier Prochaska}
 \affiliation{Department of Astronomy \& Astrophysics, University of California, 1156 High Street, Santa Cruz, CA 95064, USA}
 
\author{Joseph F. Hennawi}
\affiliation{Department of Physics, Broida Hall, University of California, Santa Barbara, CA 93106, USA}

\correspondingauthor{Ryan Rickards Vaught}
\email{rjrickar@ucsd.edu}

\begin{abstract}
We perform a Very Large Telescope FOcal Reducer and low dispersion Spectrograph 2 (VLT/FORS2) narrowband imaging search around 5 star-forming galaxies at redshift z=0.67-0.69 in the Great Observatories Origins Deep Survey South (GOODS-S) field to constrain the radial extent of large-scale outflows traced by resonantly scattered \ion{Mg}{2} emission. The sample galaxies span star formation rates in the range $4\msunperyr< \mathrm{SFR} < 40\msunperyr$ and have stellar masses $9.9 \lesssim \log M_{*}/M_{\odot} \lesssim 11.0$, and exhibit outflows traced by MgII absorption with velocities $\sim150-420\mkms$.
These observations are uniquely sensitive, reaching surface brightness limits of 5.81 $\times$ $10^{-19}$ ergs sec $^{-1}$ cm$^{-2}$ arcsec$^2$ per 1 arcsec$^2$ aperture (at 5$\sigma$ significance).  We do not detect any extended emission around any of the sample galaxies, thus placing 5$\sigma$ upper limits on the brightness of extended MgII emission of $<6.51 \times 10^{-19}$ ergs sec $^{-1}$ cm$^{-2}$ arcsec$^2$ at projected distances $R_{\perp} > 8-21$ kpc. The imaging also resolves the MgII absorption observed toward each galaxy spatially, revealing approximately constant absorption strengths across the galaxy disks. 
In concert with radiative transfer models predicting the surface brightness of MgII emission for a variety of simple wind morphologies, our detection limits suggest that either (1) the extent of the \ion{Mg}{2}-emitting material in the outflows from these galaxies is limited to $\lesssim 20$ kpc; or (2) the outflows are anisotropic and/or dusty.  
\end{abstract}

\keywords{galaxies: evolution -- galaxies: halo}

\section{Introduction}\label{sec:intro}
Galactic winds likely play a critical role in regulating the star formation rates and stellar masses of galaxies (e.g., \citethnop{SP1999}; \citethnop{Keres2009}; \citethnop{Oppenheimer2010}; \citethnop{Hopkins2014}; \citethnop{Werk_2014}); however, the physics that powers these winds remains uncertain. Some possible mechanisms have been proposed by theoretical studies that include thermal pressure from core collapse supernovae, radiation pressure from starbursts, and finally cosmic ray pressure (e.g., \citethnop{Larson_1974}; \citethnop{MO1977}; \citethnop{Chevalier_1985}; \citethnop{Breitschwerdt1991}; \citethnop{Springel_2003}; \citethnop{Murray2011}; \citethnop{Uhlig2012}).
Additionally, the impact galactic winds have on their host galaxies (i.e., their mass and energy content) has remained difficult to constrain with observations.

An accurate picture of the types of galaxies that host outflows comes from numerous absorption line studies of galaxy spectroscopy 
(e.g., \citethnop{Heckman2000}; \citethnop{Shapley2003}; \citethnop{Rupke2005b}; \citethnop{Veilleux2005}; \citethnop{Weiner2009}; \citethnop{Martin2012}; \citethnop{Rubin_2014}). Gas flows are detected by measuring the blueshift (outflow) or redshift (inflow) of absorption transitions with respect to the host galaxy systemic velocity. Spectroscopy of galaxies from low to high redshifts probing cold gas ($T \lesssim 10^2$ K) which absorbs in \ion{Na}{1} and cool gas ($T \sim 10^4$ K) absorbing \ion{Mg}{2} has revealed outflows in most galaxies that host active star formation (e.g., \citethnop{chen2010}; \citethnop{Martin2012}; \citethnop{Rubin_2014}).
However, while this technique is useful for constraining the radial velocity, column density and covering fraction of the flow, it weakly constrains the overall radial extent and provides little information on the morphology of the gas.

An alternative method that can in principle assess the radial extent and morphology of outflows is to trace the gas in emission. This has been demonstrated using rest-frame optical transitions (i.e., H$\alpha$, [\ion{O}{3}]) as tracers for winds around nearby starbursts \citep[e.g.,][]{{Heckman1990,Lehnert1999,Veilleux2003,Matsubayashi2009}} 
as these transitions are sensitive to the warm shock-heated phase of the gas. 
Another transition potentially useful for tracing winds in emission is the \ion{Mg}{2} $\lambda\lambda 2976,2803$ doublet in the rest-frame ultraviolet
\citep[UV; e.g.,][]{{Weiner2009, Kornei2013}}. While most studies of winds using \ion{Mg}{2} have focused on its absorption kinematics, \cite{Rubin_2011} observed  strong \ion{Mg}{2} emission with a P-Cygni line profile in the Keck/Low Resolution Imaging Spectrometer (LRIS) spectrum of a strongly star-forming galaxy at redshift $z = 0.694$. In addition, the emission was spatially extended beyond the galaxy continuum, permitting the first direct measurement of the extent of an outflow ($\gtrsim$ 7 kpc) in the distant universe.

One proposed production mechanism for such P-Cygni profiles is photon scattering. In this mechanism, \ion{Mg}{2} ions in the region of the wind closest to the observer will absorb continuum photons in the resonant transitions at wavelengths 2796.35\AA\ ($\lambda_{2796}$) and 2803.53\AA\ ($\lambda_{2803}$) \citep{Morton2003}. Once these transitions are excited, they may only decay back to the ground state. If the optical depth of the gas is high, then the gas will resonantly trap the absorbed photons. Because the photons are absorbed in the rest frame of the gas, the absorption is observed to be blueshifted relative to the galaxy's systemic velocity. The \ion{Mg}{2} ions in the section of the wind farthest from the observer will absorb and scatter photons that are redshifted relative to the front portion of the wind. Because the photons are redshifted, the photons travel freely toward the observer through the wind to produce emission at and redward of the systemic velocity of the galaxy (e.g., \citethnop{Rubin_2011}, \citethnop{Prochaska_2011}). 

Since the first detection of \ion{Mg}{2} emission in 
an individual galaxy by \citet{Rubin_2011}, another detection was reported by \cite{Martin2013}, who observed \ion{Mg}{2} emission that extends $12-18$ kpc from a strongly star-forming galaxy
at $z=0.9392$. \ion{Mg}{2} has also been studied in galaxy surveys conducted with Keck/LRIS, the Keck DEep Imaging Multi-Object Spectrograph (DEIMOS), the VLT Multi Unit Spectroscopic Explorer (MUSE), and the MMT Blue Channel Spectrograph  (\citethnop{Weiner2009}; \citethnop{Erb2012}; \citethnop{Kornei2013}; \citethnop{Feltre2018}; \citethnop{Henry2018}). These surveys, which include galaxies with redshifts $ 0.20 < z < 2.30$, find that \ion{Mg}{2} may be detected in pure emission, pure absorption or with P-Cygni profiles, and that detections of \ion{Mg}{2} in emission were found to be more commonly associated with galaxies of lower stellar mass and with bluer spectral slopes.

The diversity of these spectral profiles may be understood using radiative transfer modeling of galactic winds. \citet{Prochaska_2011} have used this technique to predict 
spectra for the  \ion{Mg}{2} and  \ion{Fe}{2}$^*$ fine-structure transitions for a variety of wind morphologies. The authors demonstrated that isotropic, dust-free winds will conserve photon flux, thus predicting that blueshifted absorption lines should be accompanied by emission lines with similar equivalent widths (EW). Anisotropic winds, however, were demonstrated to exhibit significantly weaker emission by a factor proportional to the angular extent (i.e., solid angle) of the wind. Scattered emission was found to be additionally weakened by the inclusion of dust and the presence of a strongly-absorbing interstellar medium (ISM). Thus, spatially-resolved measurement of the surface brightness of this emission constrains not only the radial extent of the emitting material, but also its morphology and dust content.

In this paper, we present the first narrowband imaging of the \ion{Mg}{2} transition around 5 star-forming  galaxies located in the Great Observatories Origins Deep Survey South (GOODS-S; \citethnop{Giavalisco2004})  field at redshift $z \sim 0.7$. 
We use two filters: a  ``line filter'' covering the \ion{Mg}{2} doublet in the observed frame, and a ``continuum filter'' that is offset from the line filter by ${\sim}47$ \AA.    
The resulting imaging in each filter has a total integration time of 10 hrs. As opposed to slit or fiber spectra, the narrowband imaging fully constrains the surface brightness and projected radial extent of the wind. These observations allow us to create the first ever high-S/N spatially-resolved map of both detection limits on \ion{Mg}{2} emission and on \ion{Mg}{2} absorption. 

In Section \ref{sec:obs_red} we describe our sample of GOODS-S galaxies, supplemental Keck/LRIS spectra, as well as our VLT/FORS2 observations, image reduction, and absolute flux calibration. We describe our method of continuum subtraction in Section \ref{sec.cont_sub}. Analysis of these data is presented in Section \ref{sec:analysis}, 
including our methods for calculating surface brightness profiles and detection limits for each galaxy, as well as maps of \ion{Mg}{2} equivalent widths.
Section \ref{sec:results} presents results from this analysis. We compare our surface brightness (SB) detection limits to previous detections of extended \ion{Mg}{2} emission,  and compare our observations to predictions made using radiative transfer models in Section \ref{sec:discussion}. We conclude this paper in Section \ref{sec:conclusion}.
We adopt a $\Lambda$CDM cosmology with $h_{70} = H_0/(70\ \rm{ km}\ \rm{ s}^{-1}\ \rm{ Mpc}^{-1})$, $\Omega_{\rm{M}}= 0.3$, and $\Omega_{\Lambda} = 0.7$. In this cosmology, 1\arcsec\  is $\approx 7\ \rm kpc$ at $z \sim 0.7$.

\section{Observations and Data Reduction}\label{sec:obs_red}
\subsection{Sample Selection}
\begin{figure*}[!ht]
\centering
\includegraphics[scale=.75]{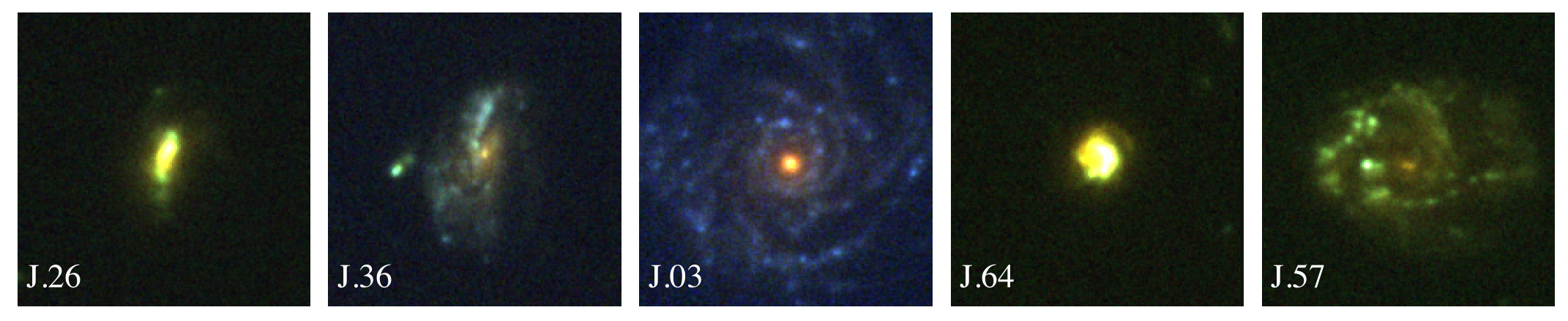}
\caption{Color imaging of our sample galaxies in the \emph{HST}/ACS F435W, F606W, and F775W filters obtained as part of the GOODS survey (\citeauthor{Giavalisco2004} \citeyear{Giavalisco2004}). Each image is 5\arcsec $\times$ 5\arcsec\ (or about $35 \rm \ kpc \times 35 \rm \ kpc$).\label{fig:hstims}}
\end{figure*}

Our target galaxies were selected from a Keck/LRIS survey of UV absorption lines in $\approx 100$ objects having redshifts $0.3< z < 1.4$ and rest-frame $B$-band magnitudes $B_{\rm AB}< 23$ in fields with deep Hubble Space Telescope/Advanced Camera for Surveys (\emph{HST}/ACS) imaging \citep{Rubin_2014}.  In particular, this parent survey targeted galaxies in a total of nine Keck/LRIS pointings located in both of the GOODS fields (\citeauthor{Giavalisco2004} \citeyear{Giavalisco2004}) and the AEGIS survey field \citep[the Extended Groth Strip;][]{Davis2007}.  In inspecting the redshift distribution of the portion of this sample observable from the Southern Hemisphere, we uncovered a narrow peak of nine galaxies in the interval $0.66 \lesssim z \lesssim 0.68$.  This peak is in fact the global maximum of the distribution, as all other bins of width $\Delta z = 0.02$ have at most four galaxies.  Moreover, there are two narrow interference filters available on VLT/FORS2 centered at $\lambda \sim 4675$ \AA\ and 4722 \AA\ which cover the \ion{Mg}{2} $\lambda \lambda 2796, 2803$ transition in precisely this redshift interval.  We selected our final sample of five of these galaxies 
to be close on the sky such that they could be imaged in a single $7' \times 7' $ FORS2 pointing.  
We show color {\it HST}/ACS images of these objects in Figure~\ref{fig:hstims}.

The Bayesian absorption line modeling presented in \cite{Rubin_2014} indicates that these five galaxies are driving strong outflows traced by \ion{Mg}{2} with maximum outflow velocities $\Delta v_{\rm max}$ $\sim150-420\mkms$ and rest-frame equivalent widths $\sim 2-3$ \AA. These maximum outflow velocities (listed in Table \ref{tab:prop}) are determined from fitting a two-velocity component model to the absorption line profiles. The two-component model assumes that there is an absorption component due to stellar atmospheres and the interstellar medium with a velocity fixed at systemic, as well as a ``flow'' absorption component with a velocity that is allowed to float.  
The maximum outflow velocity $\Delta v_{\rm max}$ is the fitted central velocity of the flow component, $v_{\rm flow}$, minus the fitted Doppler parameter, $b_{D,\rm flow}/\sqrt{2}$. It is thus indicative of the most extreme flow velocities traced by each absorption line profile.
The two-component model does not explicitly include a contribution to the line profile from scattered emission, which
can be significant at velocities close to systemic \citep{Prochaska_2011}.  However, as discussed in \cite{Rubin_2014}, it is expected that scattered emission will primarily tend to reduce the strength of the fitted systemic component, and will have a minor effect on fitted flow component velocities and line widths (see their Appendix C).

Modeling of the galaxy broad-band spectral energy distributions (SEDs) obtained from multi-wavelength ancillary imaging data, also performed by \cite{Rubin_2014}, yields star formation rates (SFR) ranging from $\sim4$ to $40\msunperyr$ and stellar masses in the range $\log M_*/M_{\odot}\sim 9.9-11.0$. All of these sample properties as well as target coordinates 
are listed in Table \ref{tab:prop}.

\subsection{VLT/FORS2 Observations}
\begin{figure*}[!h]
\centering
\gridline{\fig{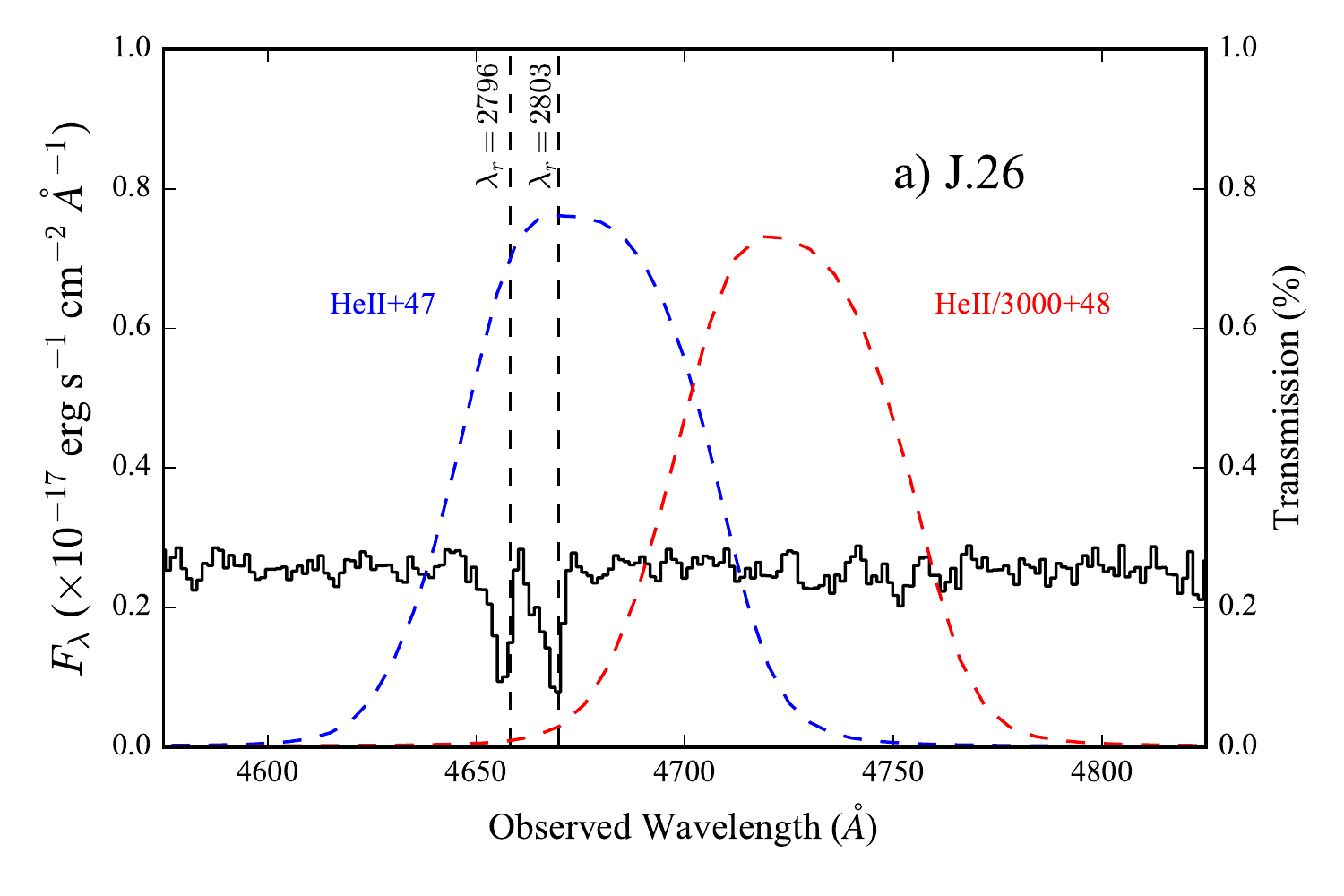}{0.5\textwidth}{}
          \fig{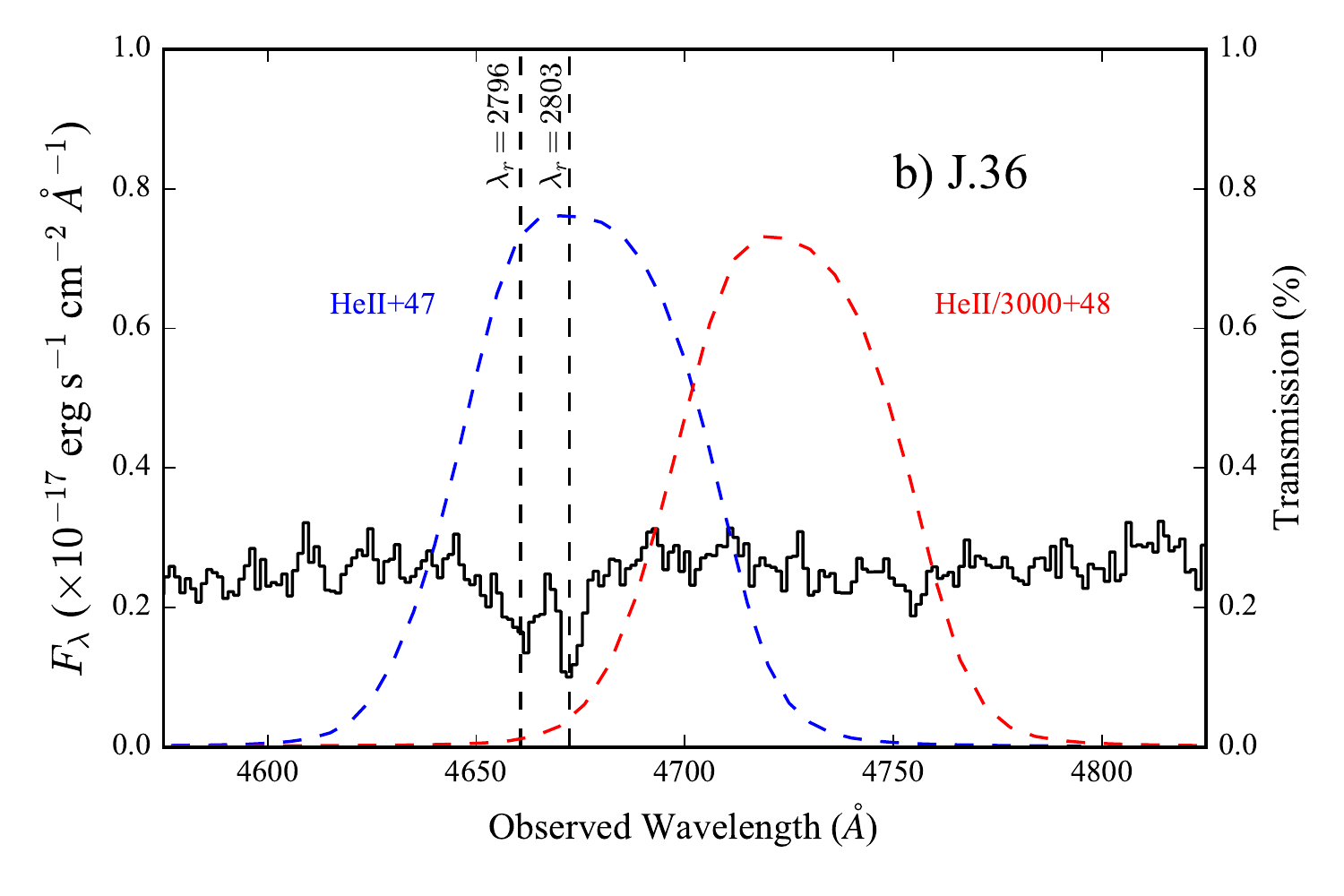}{0.5\textwidth}{}}
\gridline{\fig{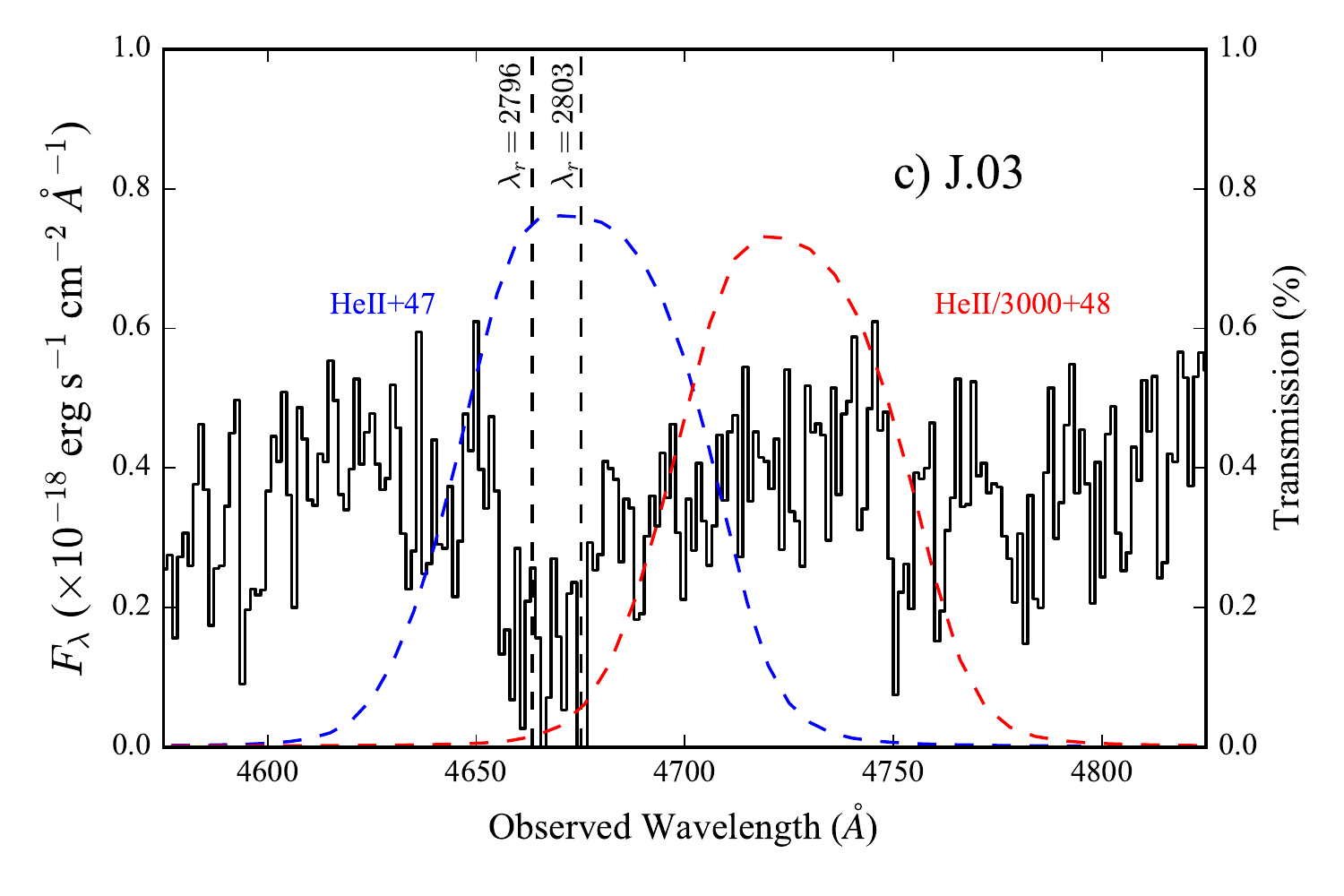}{0.5\textwidth}{}
          \fig{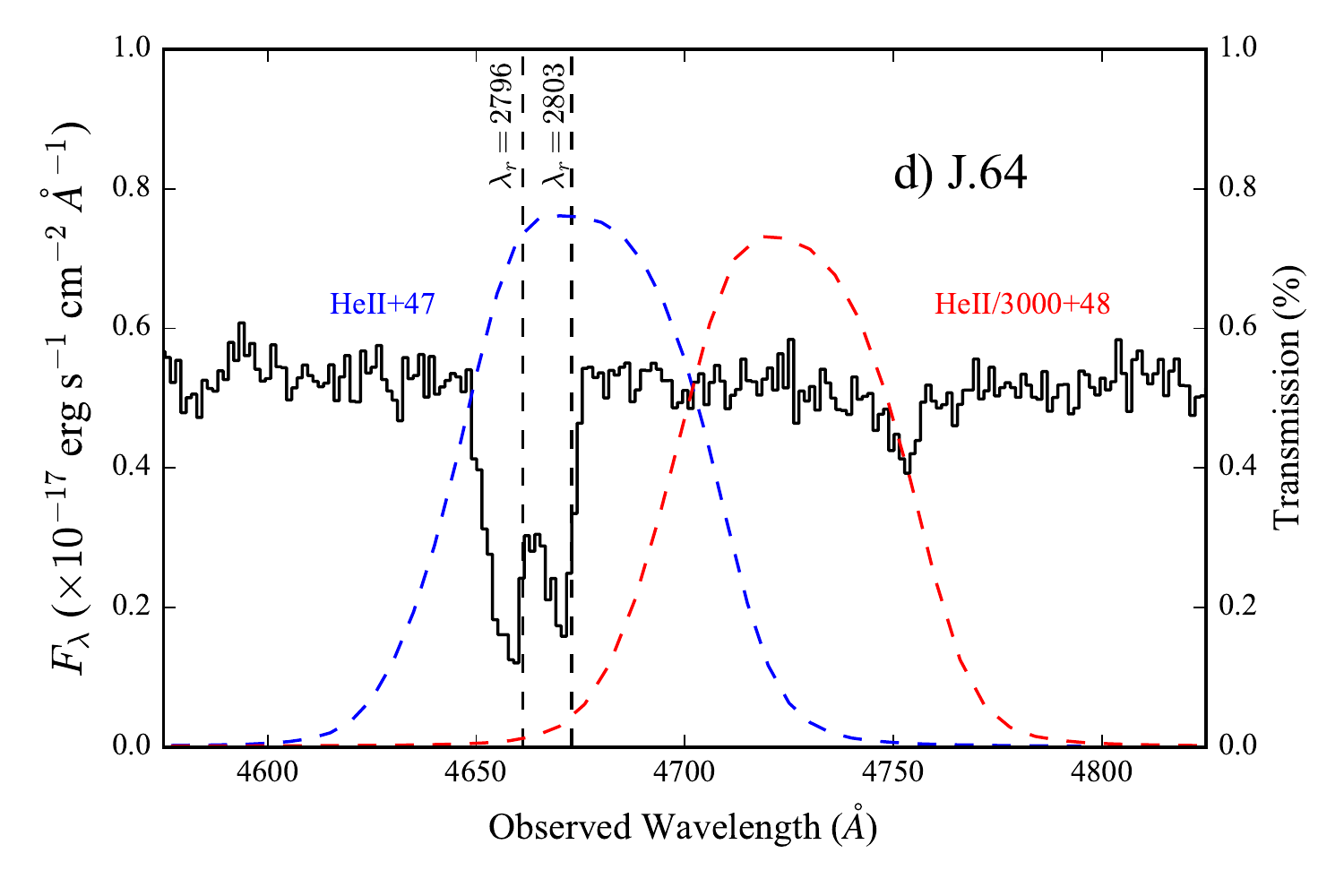}{0.5\textwidth}{}}
         \fig{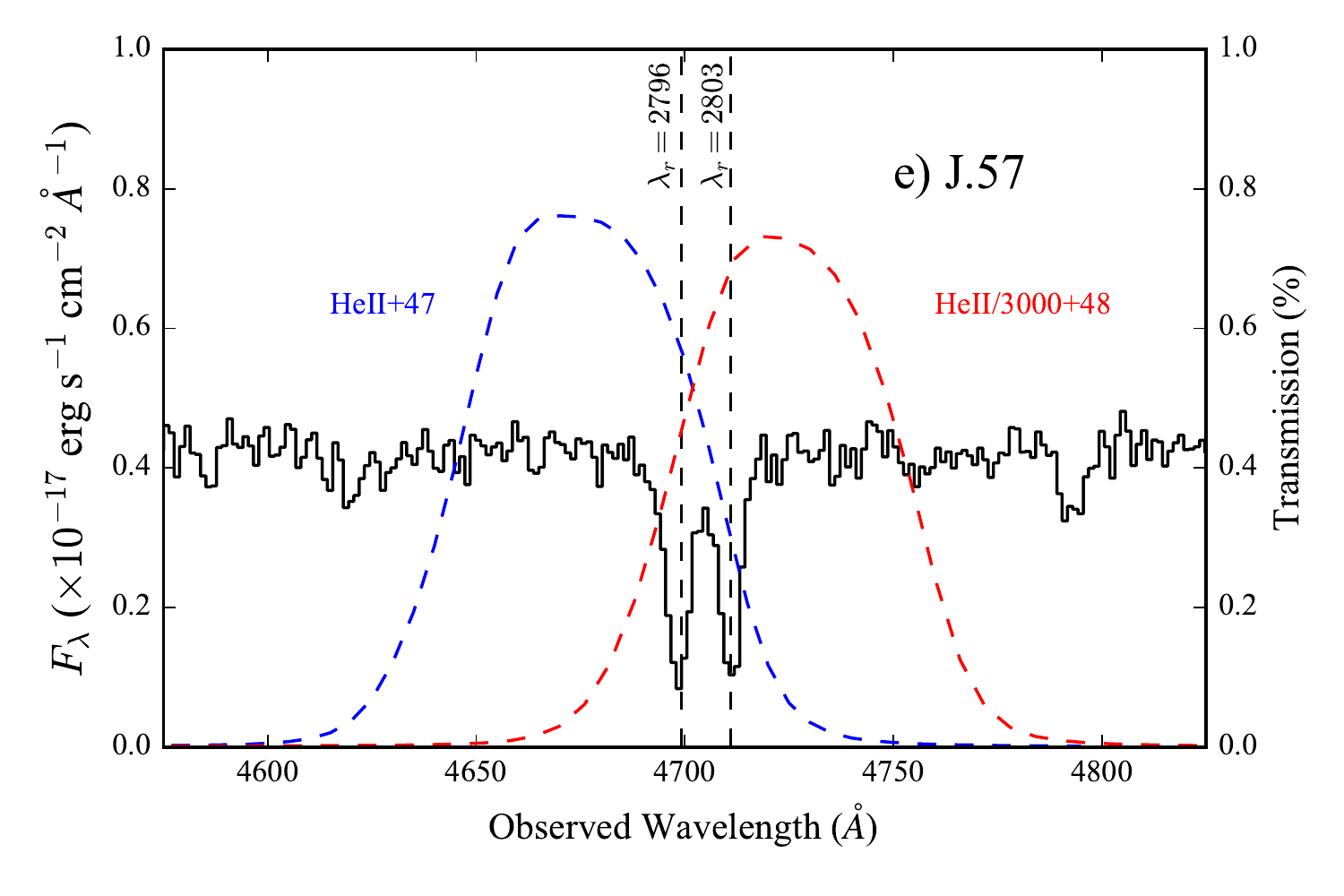}{0.5\textwidth}{}
\caption{Keck/LRIS spectra of the sample galaxies and the transmission curves of the filters HeII+47 (blue dashed line) and HeII/3000+48 (red dashed line). The left-hand axis is in units of flux density and the right-hand axis is the percentage of light transmitted by the filter at each wavelength. Vertical dashed lines indicate the wavelengths of the redshifted \ion{Mg}{2}\ doublet. The \ion{Mg}{2}\ doublet falls fortuitously at the central wavelength of the HeII+47 filter for the galaxies shown in panels (a) through (d).}
\label{fig:spec_images}
\end{figure*}

Our narrowband imaging data were taken in service mode using the FORS2 instrument on the VLT 8.2m telescope Antu between October 2012 and February 2013. 
We used two narrowband filters, HeII+47 and HeII/3000+48, that have peak transmission at wavelengths that correspond to the \ion{Mg}{2} doublet lines at our sample redshift of $z\sim0.7$ (see Table \ref{tab:filters}). The filter transmission curves are plotted along with each galaxy's spectrum in Figure~\ref{fig:spec_images}.
In the following, we will often refer to the HeII+47 filter as the ``line'' or \ion{Mg}{2} filter and the HeII/3000+48 filter as the ``continuum'' filter.

\begin{figure}[h]
\centering
\includegraphics[scale=.55]{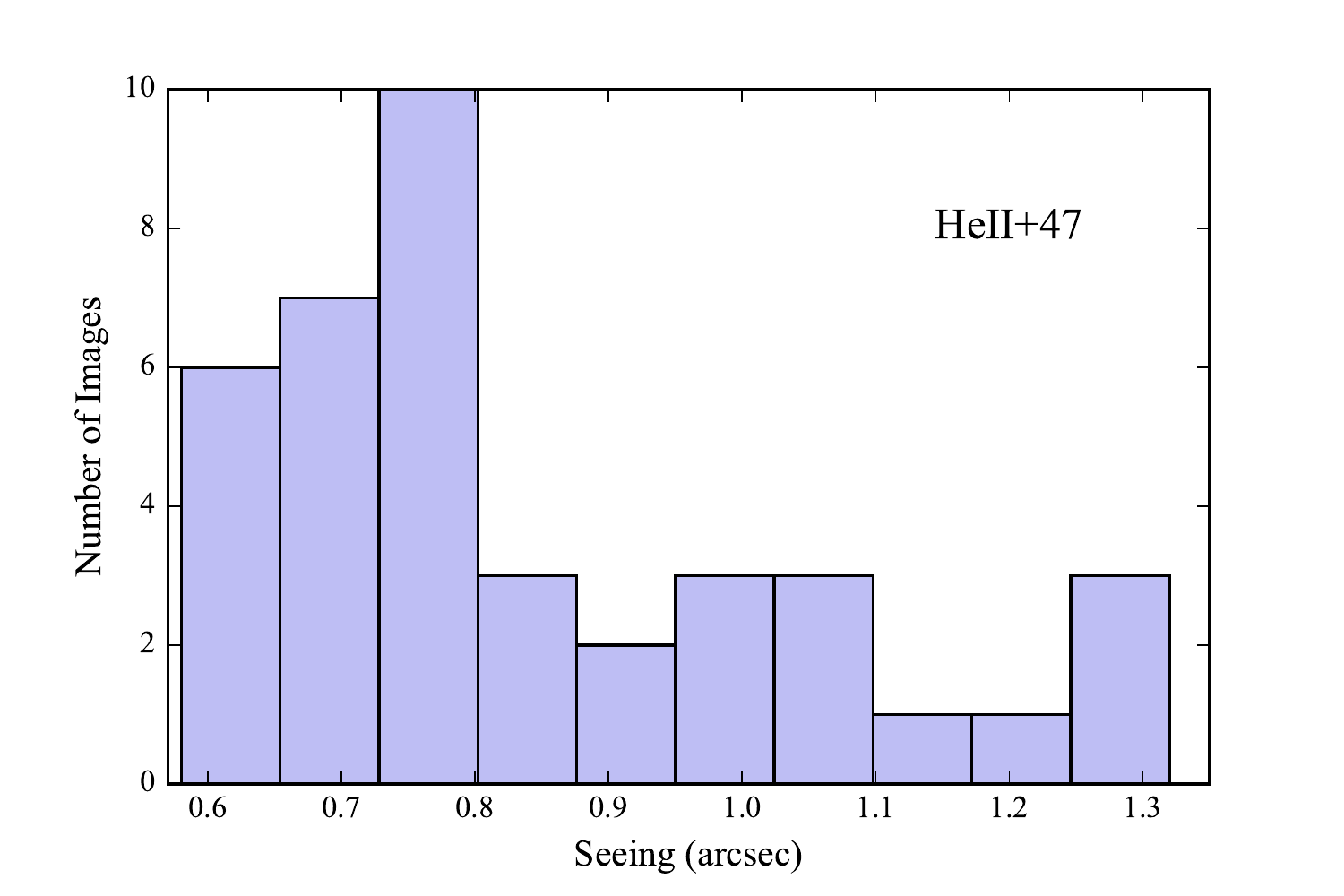}
\includegraphics[scale=.55]{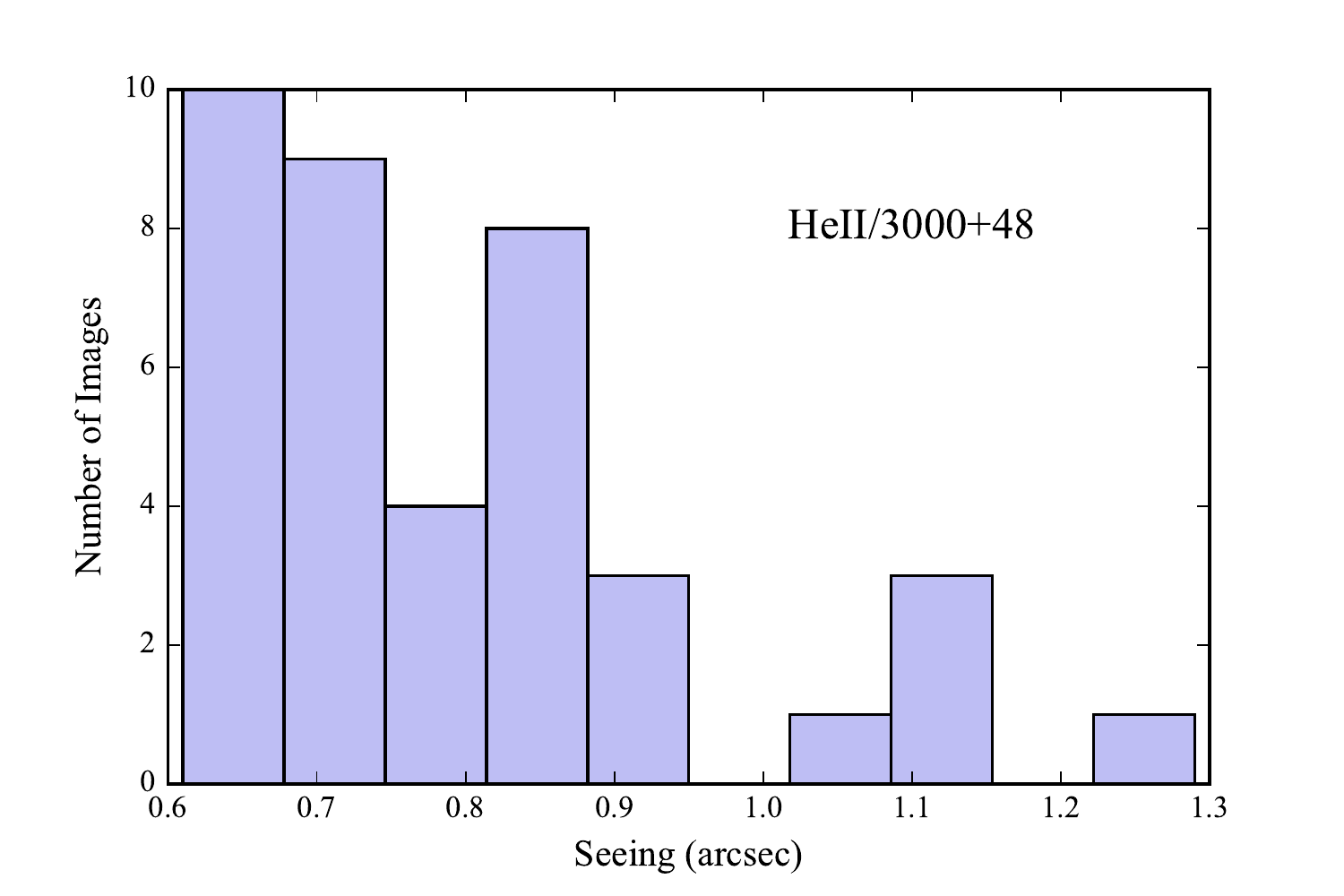}
\caption{\emph{Top}: Distribution of seeing measurements for the 38 HeII+47 images.
\emph{Bottom}: Same for the HeII/3000+48 images. The median seeing value for the images in both filters is $\sim 0.8\arcsec$. 
The seeing conditions were calculated from DIMM measurements which are provided by ESO in the header of each science image.
\label{fig.seeing}}
\end{figure}

FORS2 has a native pixel scale of $0.125''$ pixel$^{-1}$ and a field of view of $7'\times7'$.  The data were taken with 
the CCD binned $2\times2$, yielding a pixel scale of $0.25''$ pixel$^{-1}$.
Images of three pointings offset by $0.25\arcmin$ East/West were obtained, with individual exposure times of $\approx$ 1000 sec.  A total of 38  exposures were taken in each filter. 
Our observations were carried out under photometric and thin cloud conditions (program ID: 090.A-0427A). 
The seeing values, given in the header of each image, were derived from zenith observations at 0.5 micron with the Paranal differential image motion monitor \citep[DIMM;][]{Sarazin1990} and include a correction for the airmass and wavelengths of the science observations, as well as a first order correction for the larger size of the Antu mirror. The distribution of these seeing values is shown in Figure \ref{fig.seeing}. The median seeing for the images is $\sim 0.8\arcsec$. Summing the individual exposure times for each filter results in a combined exposure time of $10.0$ hours each for the HeII+47 and HeII/3000+48 images.

\subsection{Supplemental Keck/LRIS Spectra}
In addition to VLT imaging, in the present analysis we utilize galaxy spectra taken from the \cite{Rubin_2014} Keck/LRIS program.  A $0.9''$ slit width was used for all slitmasks and the median FWHM resolution for the spectra is 274 km s$^{-1}$ at $\lambda_{\rm rest} \approx$ 2800 \AA\ and $286\mkms$  at $\lambda_{\rm rest}\approx$ 2600 \AA\ (see Figure \ref{fig:spec_images}).  The spectral coverage of these data extends from ${\sim}3200$ to 8000 \AA.

\subsection{Image Reduction}
The imaging data were fully reduced using custom routines written in \emph{Python}. 
The images were first corrected by subtracting and removing the overscan region of the CCD. 
Then the images were bias-subtracted and flat-fielded using twilight flats.
An additional flat-fielding correction was performed using night-sky flats to improve our sensitivity to faint extended emission.  The night-sky flats were produced by first masking out all objects and bad pixels in the science frames, and then combining them using an average sigma-clipping algorithm.
Cosmic rays and bad pixels in the science images are removed by utilizing the \emph{L. A. Cosmic} algorithm \citep{Dokkum2001}.
The astrometry solutions were calculated via Astrometry.net \citep{Lang}, and yield a standard deviation in the galaxy coordinates of $\sigma \approx 0.10''$. Before image stacking, we ran each frame through \emph{SExtractor} \citeth{Bertin} to create a root mean square (RMS) map of each science image.

\begin{figure*}[ht!]
\centering
\includegraphics[scale=.61]{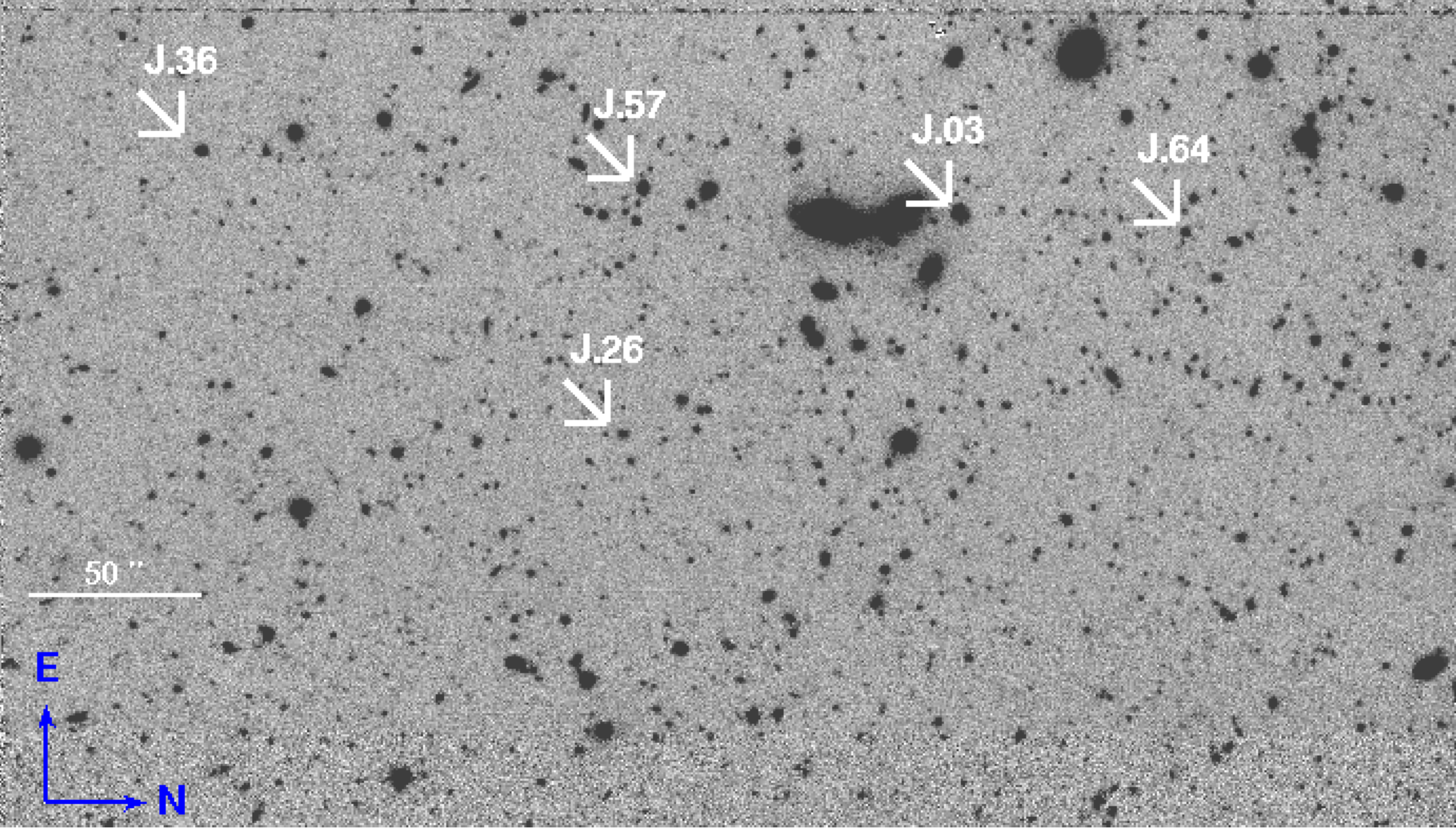}
\includegraphics[scale=.61]{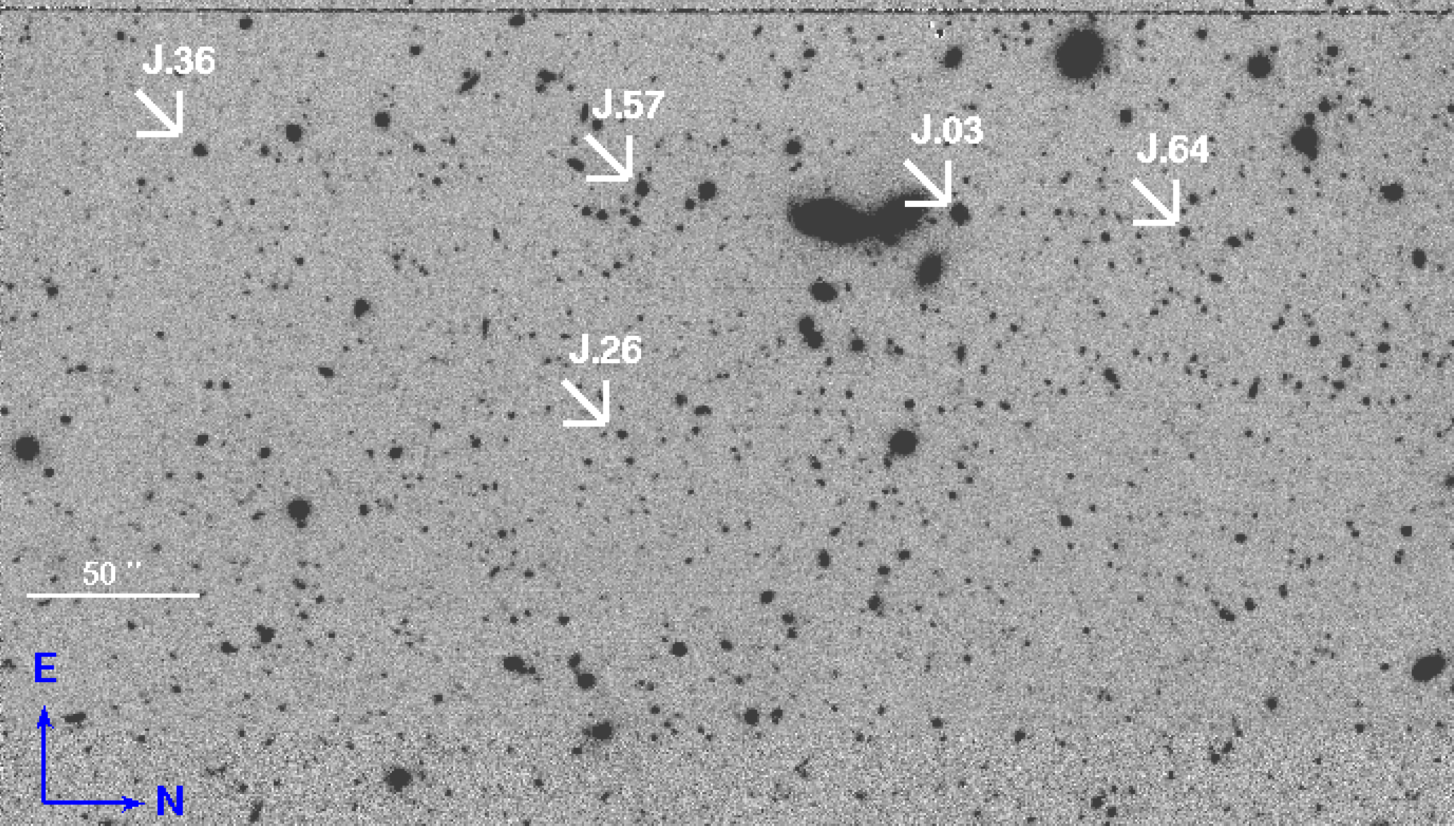}
\caption{\emph{Top:} Stacked HeII+47 image of the galaxy sample. \emph{Bottom:} Stacked HeII3000+48 image of the same pointing. The exposure time of each image is $\approx 10$ hours. Each image shows approximately half of the total FORS2 FOV ($7' \times 5'$), which contains the full sample of galaxies (indicated by the white arrows). East is up and North is right.
\label{fig:stacked_image}}
\end{figure*}

The final stacked image for each filter is obtained using \emph{SWarp} \citeth{Bertin}.
Each individual frame is first sky-subtracted using a background mesh size of 256 pixels which is approximately $64''$. 
We chose the mesh size to be large enough such that any extended emission is not mistakenly subtracted \citep[e.g.,][]{Battaia_2015}. 
The frames, after background-subtraction, are resampled onto a common astrometric solution using a \textit{Lancosz3} interpolation kernel. 
The images are weighted by the night-flat image and then  average-combined to increase the signal-to-noise of any \ion{Mg}{2} emission. Additionally, \emph{SWarp} generates stacked RMS images by propagating the error images for each science frame.
Our final stacked images in each filter are shown in Figure~\ref{fig:stacked_image} with the target galaxies indicated.

\begin{deluxetable*}{c c D D D c D}
\tablecaption{Properties of the 5 galaxies in our sample as estimated in \cite{Rubin_2014}. The EW (in the observed frame) includes both components of the \ion{Mg}{2} doublet and is 
determined from analysis of the supplemental Keck/LRIS spectra. \label{tab:prop}}
\tablehead{\colhead{Object\tablenotemark{a}} &  \colhead{$z$} & \multicolumn2c{SFR} & \multicolumn2c{$\log{(M)}$} &\colhead{$\Delta v_{\rm max}$\tablenotemark{b}} &\multicolumn2c{EW$_{\rm{obs}}$} & \multicolumn2c{ $\tau_V$\tablenotemark{c}} \\ 
\colhead {} & \colhead{} & \multicolumn2c{($M_{\odot}$ yr$^{-1}$)} & \multicolumn2c{($\log{M_{*}/M_{\odot}}$)} & \colhead{(km s$^{-1}$)} & \multicolumn2c{(\AA)} & \multicolumn2c{}}
\decimals
\startdata
J033225.26-274524.0 (J.26)     &   0.6660  & $9.1_{-3.7}^{+1.3}$     & $9.86_{-0.04}^{+0.05}$   &   $-187_{-16}^{+12}$ & \ $7.5\pm 0.4$   &  $1.227_{-0.20}^{+1.54}$ \\ 
J033229.64-274242.6 (J.64)     &   0.6671  & $40.5_{-12.1}^{+8.2}$ & $10.30_{-0.03}^{+0.07}$ &   $-378_{-12}^{+12}$ & $13.2 \pm 0.3$ & $3.897_{-0.93}^{+1.15}$ \\
J033230.03-274347.3 (J.03)     &   0.6679  & $3.8_{-0}^{+0}$           & $10.98_{-0.0}^{+0.01}$   &   $-400_{-64}^{+132}$ & $12.8 \pm 1.7$ & $0.297_{-0.0}^{+0.0}$ \\
J033230.57-274518.2 (J.57)     &   0.6807  & $12.6_{-2.1}^{+1.7}$   & $10.48_{-0.07}^{+0.03}$ &   $-266_{-38}^{+44}$ & \ $6.1 \pm 0.4$   & $1.262_{-0.40}^{+1.23}$ \\
J033231.36-274725.0 (J.36)     &   0.6669  & $10.5_{-1.6}^{+1.7}$   & $10.02_{-0.03}^{+0.03}$ &   $-168_{-126}^{+33}$ & \ $5.8 \pm 0.5$   & $1.377_{-0.23}^{+0.60}$ \\
\enddata
\tablenotetext{a}{Galaxy names include their R.A. and Declination in the J2000.0 epoch.  The names in parentheses are used to identify each object throughout the paper.}
\tablenotetext{b}{Maximum outflow velocity traced by \ion{Mg}{2}, $\Delta v_{\rm max} \approx v_{\rm flow}- b_{D,\rm flow}/\sqrt{2}$, where $b_{D,\rm flow}$ is the fitted Doppler parameter and $v_{\rm flow}$ is the fitted central velocity of the ``flow'' component in a two-component model of the absorption line profile.}
\tablenotetext{c}{Total $V$-band optical depth of dust attenuating light from the young stellar population in each galaxy as modeled by MAGPHYS.  This includes contributions from dust in both \ion{H}{2}  regions and the ambient ISM.} 
\tablecomments{The uncertainty intervals reported for SFR, $\log M_*/M_{\odot}$, and $\tau_V$ values indicate the $\pm34$th-percentile values of the posterior probability distribution function of each quantity as determined via modeling of the objects' SEDs with MAGPHYS (\citeauthor{daCunha2008} \citeyear{daCunha2008}, \citeyear{daCunha2012}) as described in \citet{Rubin_2014}.}
\end{deluxetable*}

\begin{deluxetable}{c c c c c c}
\tablecaption{Filter properties  and exposure times of the VLT/FORS2 observations. The widths of the transmission curves ($\Delta\lambda$) are calculated by convolving the transmission curves over the total wavelength range of each filter. 
 \label{tab:filters}}
\tablehead{\colhead {Filter (Line)} & \colhead {$\lambda_{\rm{eff}}$\tablenotemark{a}} & \colhead {$\Delta\lambda$\tablenotemark{b}} & \colhead {N\tablenotemark{c}} & \colhead{$T$\tablenotemark{d}} & \colhead{$S$\tablenotemark{e}} \\
\colhead {} & \colhead{(\AA)} & \colhead{(\AA)} & \colhead{} & \colhead{(sec)} & \colhead{}}
\startdata
HeII+47 (\ion{Mg}{2})         & 4675.21  & 50.11 & 38  & 35,959 & 2.45 \\
HeII/3000+48 (Cont.) & 4722.46  & 44.82 & 38  & 36,937 & 2.40  \\
\enddata
\tablenotetext{a}{$ \lambda_{\rm{eff}}$ is the effective wavelength of the filter transmission curve.}
\tablenotetext{b}{The effective width of the filter.}
\tablenotetext{c}{Total number of images.}
\tablenotetext{d}{Total exposure time.}
\tablenotetext{e}{S, the sensitivity of the filter, is in units of $10^{-17}$ ergs counts$^{-1}$ cm$^{-2}$.}
\end{deluxetable}\

\subsection{Absolute Flux Calibration}
We acquired observations of the standard star GD50 from archival European Southern Observatory (ESO) calibration imaging at 4 independent epochs. Performing aperture photometry at each epoch and airmass, we calculated the atmospheric extinction coefficients, $k$, to be 0.181 magnitudes for the HeII/3000+48 filter and 0.190 magnitudes for the HeII+47 filter. 
We perform absolute flux calibration using the methods of \cite{Jacoby1987}. We first convolve the spectral energy distribution of the standard star, $F(\lambda)$ in ergs sec$^{-1}$ \AA$^{-1}$ cm$^{-2}$, with that of the known transmission curve of the filter, $T_{i}(\lambda)$. This yields $F_i$, the total observable flux in each bandpass filter $i$ with units of ergs sec$^{-1}$ cm$^{-2}$:
\begin{equation*}
F_{i}=\int F(\lambda)T_{i}(\lambda)d\lambda.
\end{equation*}
It is not uncommon to assume that $F(\lambda)$ is constant over the small width of the filter. 
However, since our filter transmission curves are sampled at wavelength intervals similar to the sampling of the spectrum of GD50 from the latest CALSPEC spectral library \citep{Bohlin2017}, we interpolate both spectra and compute the integral without the above assumption. 
The conversion from count rate to flux units for each filter is then given by
\begin{equation*}
S_{i}=\dfrac{F_{i}}{C10^{k_{i}A}},
\end{equation*}\\
where $k_i$ is the extinction in magnitudes per airmass, A is the airmass for each individual exposure, C is the measured count rate of the standard star and $S_i$ is in units of ergs counts$^{-1}$ cm$^{-2}$. Before image co-addition, each science image is corrected for atmospheric extinction by multiplying each frame by $10^{k_{i}A}$. Next, the image is divided by the exposure time, effectively putting the image in units of counts per sec. After co-addition, the images are then multiplied by the appropriate sensitivity factor $S_{i}$. This puts the final images in the appropriate flux units, ergs sec $^{-1}$ cm$^{-2}$.

\section{Image Subtraction}\label{sec.cont_sub}
We have two goals for our study: (1) assess the surface brightness of line emission in the \ion{Mg}{2} transition in and around each target galaxy; and (2) spatially resolve the morphology of the strong \ion{Mg}{2} absorption observed against the galaxy continua.
To achieve both of these goals, we must perform accurate subtraction of the continuum flux of each object from the images taken with the filter covering the targeted line emission. For four of the five galaxies in our sample, the HeII+47 image includes both line and continuum emission, and the HeII3000+48 image provides a high S/N measurement of the continuum only $\approx30$ \AA\ redward of the line emission in the rest frame. 
The spectral coverage of these filters is qualitatively different for the fifth galaxy in our sample (J.57).
As shown in Figure \ref{fig:spec_images}, the \ion{Mg}{2} transitions in this galaxy are approximately equally sampled by both of our filters. When we subtract the continuum image from the \ion{Mg}{2} image we are effectively subtracting both \ion{Mg}{2} emission (if present) and the continuum. We thus use this galaxy as check on the quality of our continuum subtraction.

\subsection{Spectral Correction}
In preparation for continuum subtraction, we first consider whether the continuum level of each galaxy spectrum changes significantly over the passbands of our two filters.
We use the supplementary spectra from \citet{Rubin_2014} to fit the continuum and determine the spectral slope of each galaxy. We use the interactive fitting routine \emph{lt\_continuumfit} from the \emph{linetools} package \citep{Prochaska2016}\footnote{https://github.com/linetools/linetools} to fit the continuum. We then find the total continuum flux in each filter by convolving the fitted continuum with each filter's transmission curve. Next, we take the ratio of both integrated totals, as the ratio will indicate the scaling factor needed to correct our flux measurements prior to continuum subtraction. 

Comparing these ratios between each galaxy, we find that they are equivalent to within 0.1\%, with a value of 1.118. This value is equal to the ratio between the effective widths of the filter transmission curves, indicating that the spectral slope of each galaxy is approximately flat, and that the continuum level measured in the off-line filter provides an accurate measure of the continuum contribution to the on-line filter flux.  We thus do not apply any spectral correction in the following analysis.

\subsection{Continuum Subtraction}\label{subsec.cont_sub}

\begin{figure*}[!htb]
\centering
\includegraphics[scale=0.7]{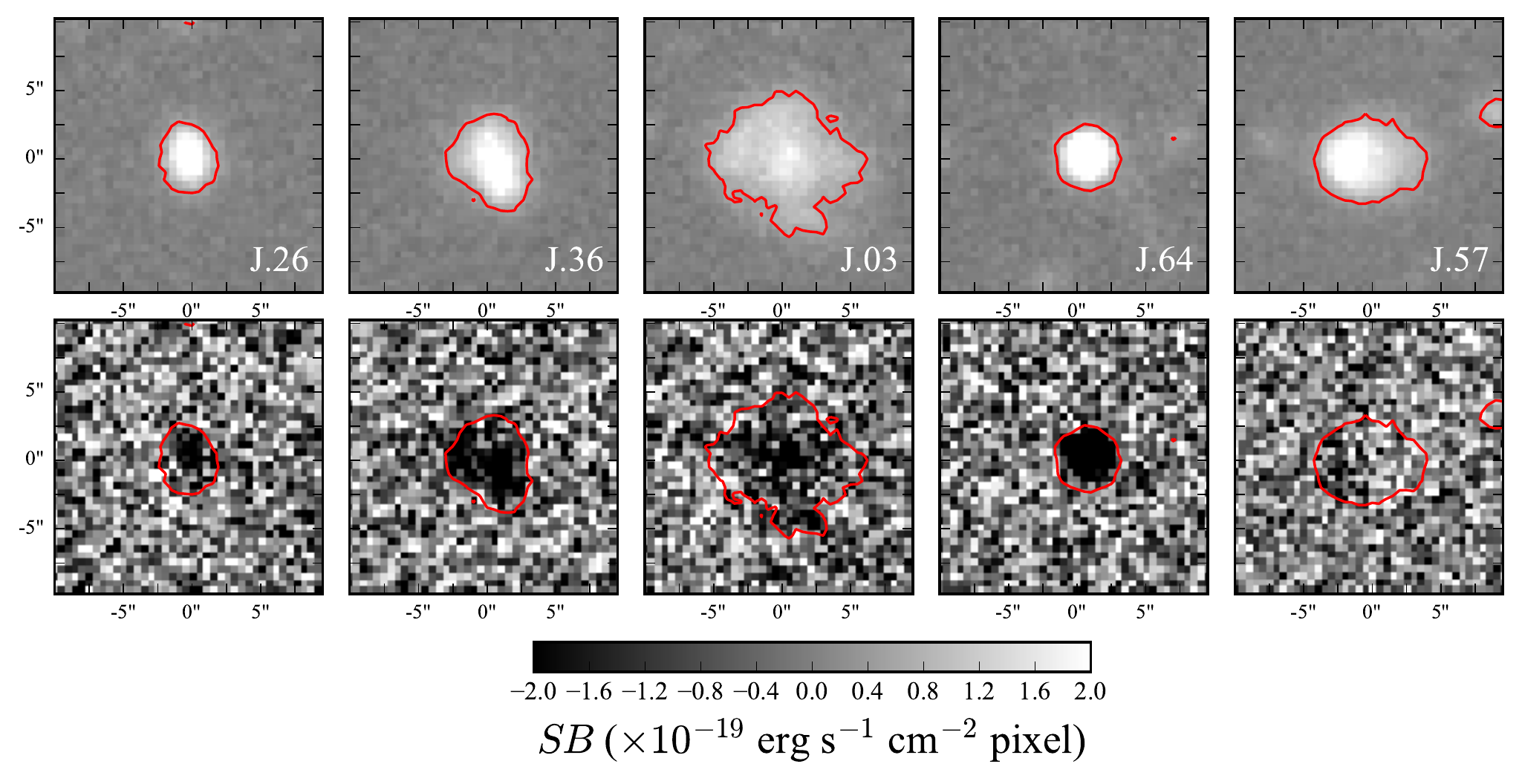}
\caption{ $10'' \times 10''$ (or about $70 \rm \ kpc \times 70 \rm \ kpc $) images of each galaxy in our sample. Top row: Continuum surface brightness in ergs s$^{-1}$ cm$^{-2}$ pixel, measured in the HeII/3000+48 filter. Bottom row: Continuum-subtracted \ion{Mg}{2} surface brightness.  Absorption can be seen in 4 of 5 galaxies. The red contours represent the outline of the 1$\sigma$ surface brightness limit in the HeII+47 image, defined in Sec. \ref{sec.sb}. The colorbar shows the scaling used for the \ion{Mg}{2} images in the bottom row.}
\label{fig:stamp_images}
\end{figure*}

To properly continuum-subtract the image taken with the \ion{Mg}{2} filter, we follow a prescription given by \cite{Battaia_2015}. 
We first determine the continuum flux density from the continuum filter,
\begin{equation}
f_{\rm{cont}}=\frac{F_{\rm{cont}}}{\Delta \lambda_{\rm{cont}}},
\end{equation}\\
where $F_{\rm{cont}}$ and $\Delta \lambda_{\rm{cont}}$ are the observed flux per pixel of the continuum image and the effective width of the continuum filter, respectively. With $f_{\rm cont}$ it is then possible to calculate the flux of any excess emission, $F_{\rm{line}}$:
\begin{equation}
F_{\rm{line}}=F_{\rm{MgII}}-f_{\rm{cont}} \Delta \lambda_{\rm{MgII}}
\label{eq:subtraction}
\end{equation}
where $F_{\rm{MgII}}$ and $\Delta \lambda_{\rm MgII}$ are the observed flux per pixel in the \ion{Mg}{2} filter and the effective width of the \ion{Mg}{2} filter. The continuum-subtracted images of each galaxy are shown in Figure \ref{fig:stamp_images}. 
These
images have a uniform background and no obvious signatures of emission.

\section{Analysis} \label{sec:analysis}

\subsection{Surface Brightness Profiles and Limits}\label{sec.sb}
\begin{figure*}
\centering
\gridline{\fig{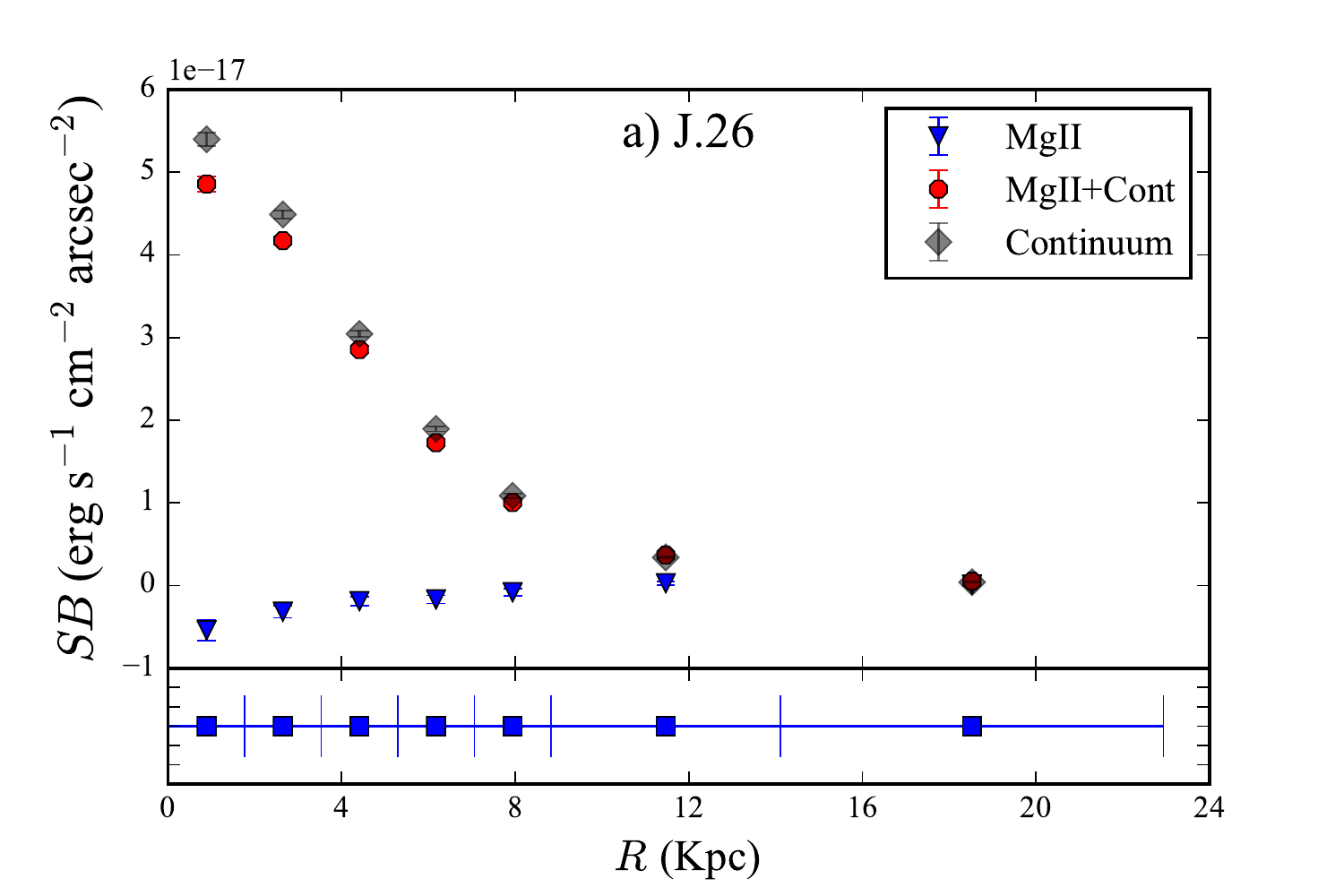}{0.5\textwidth}{}
          \fig{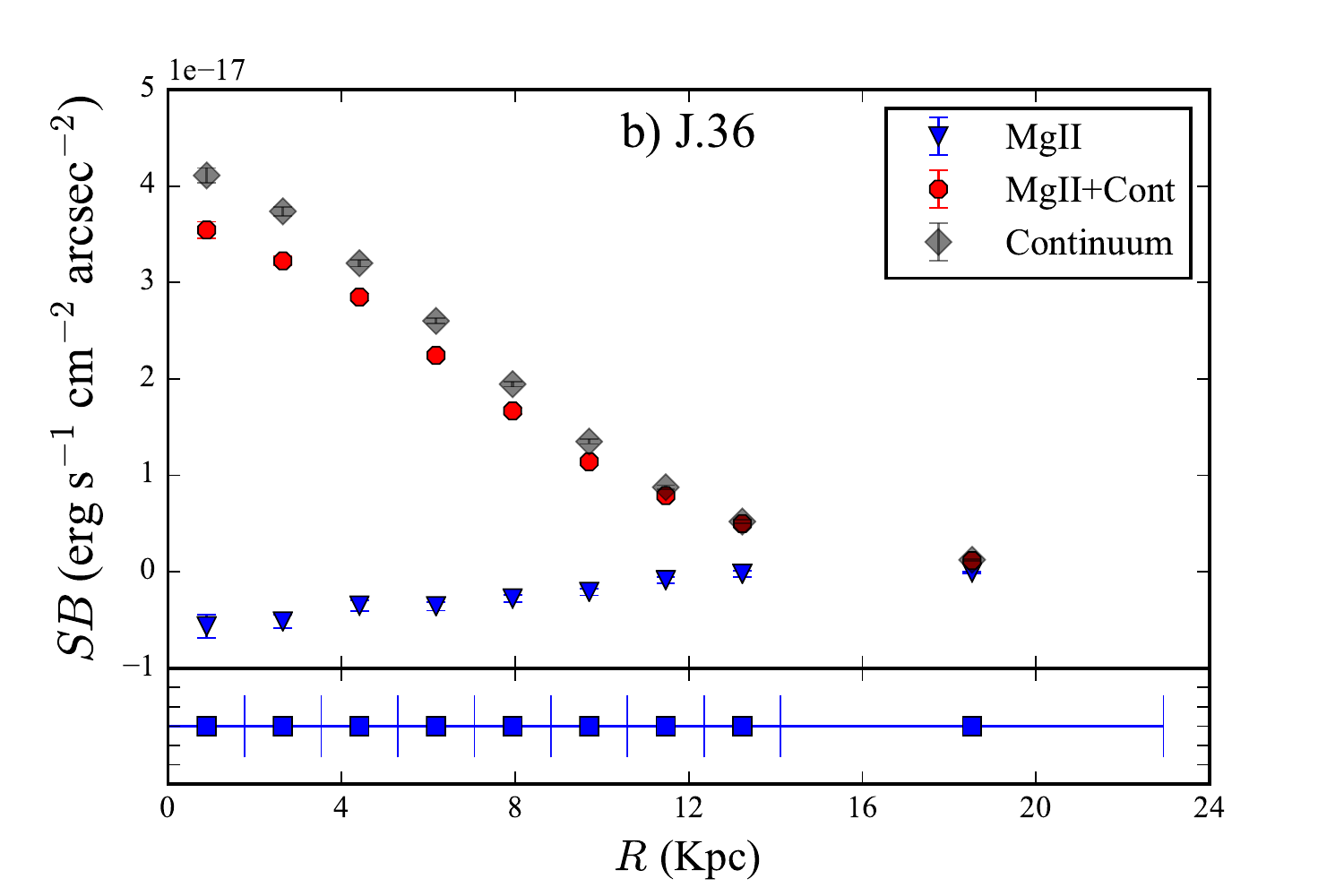}{0.5\textwidth}{}}
\gridline{\fig{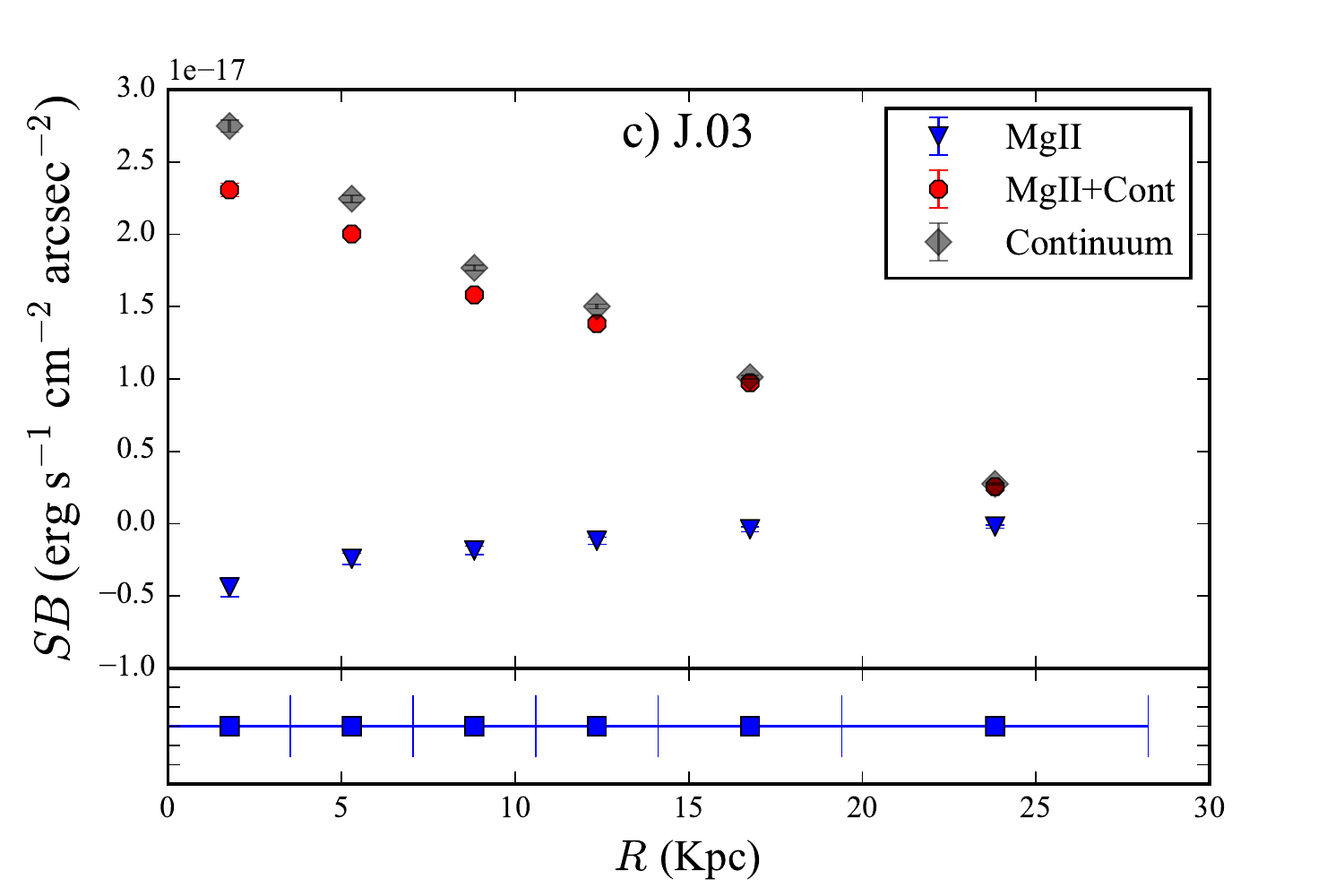}{0.5\textwidth}{}
          \fig{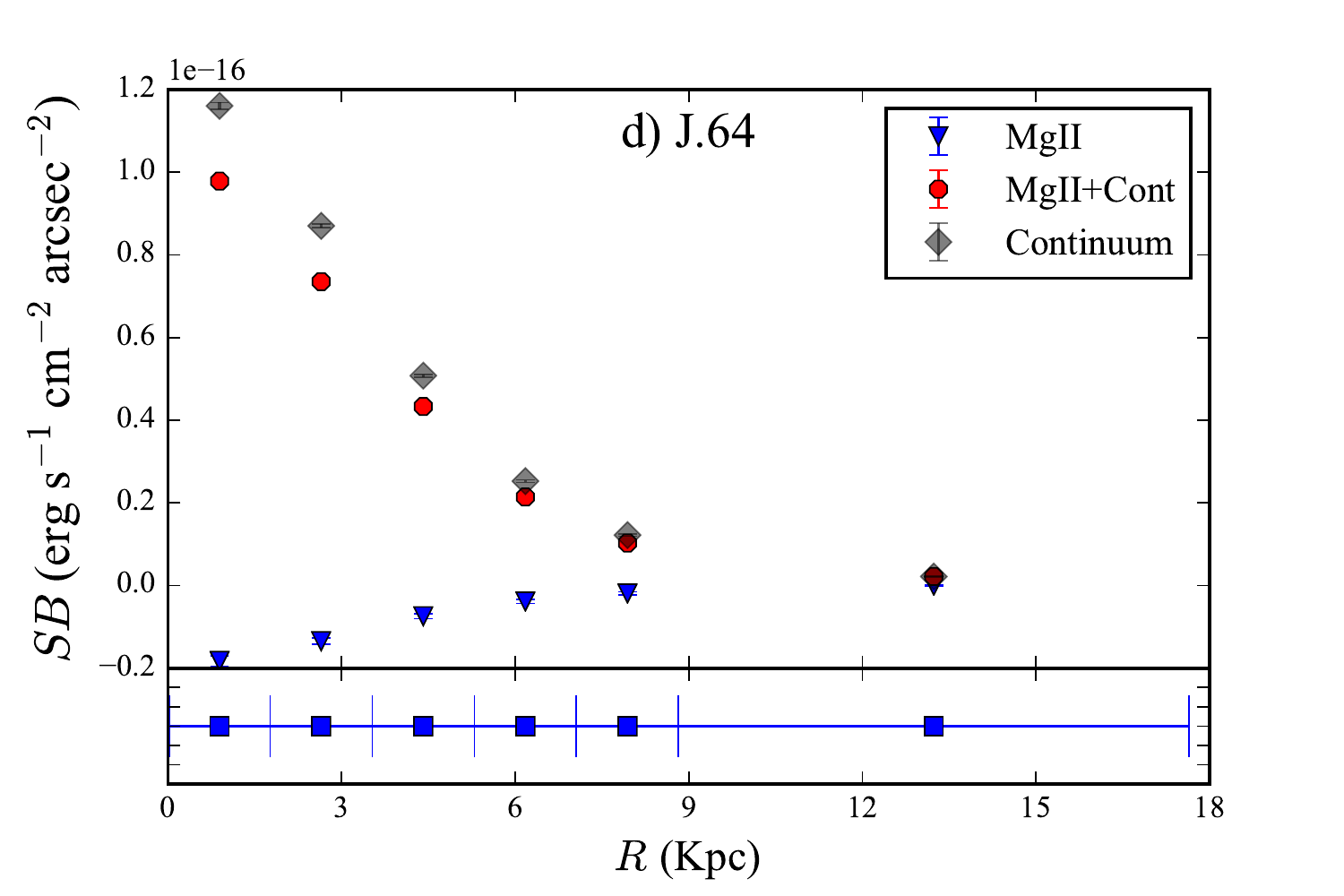}{0.5\textwidth}{}}
           \fig{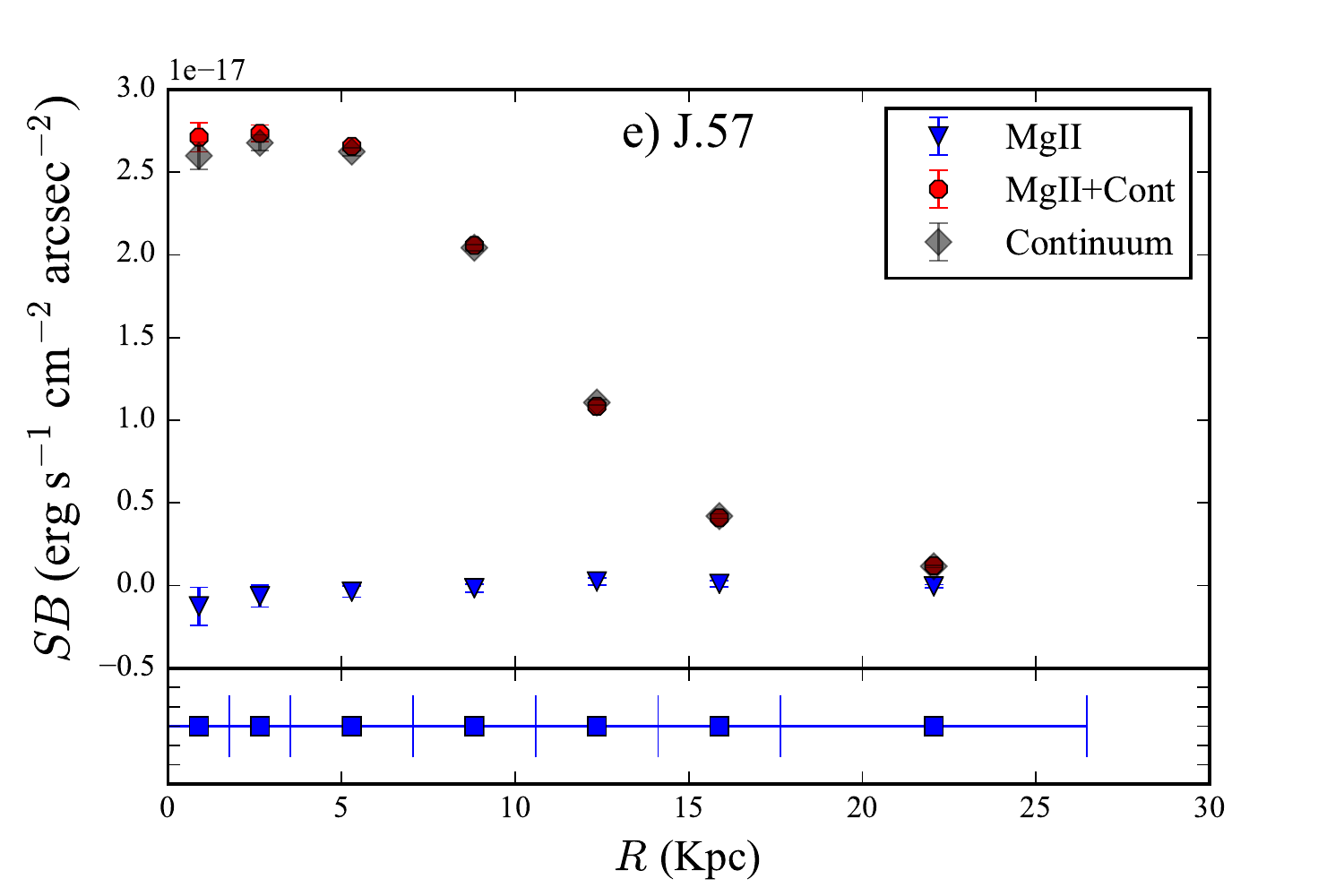}{0.5\textwidth}{}
\caption{SB profiles for our sample galaxies. Top panels: Continuum SB profile (gray) measured for each galaxy. The red points show the Mg II + continuum SB measured for the galaxy in the pre-continuum-subtracted line image. The blue points show the Mg II line SB measured  in the continuum-subtracted line emission image.  The profile exhibits SB decrements from \ion{Mg}{2} absorption. Photometry was performed in circular annuli. Bottom panel: The vertical hashes show the inner and outer radii of each annulus in kpc. The distance from the center of the galaxy is marked with filled squares and is computed using the average value of the inner and outer radii of each annulus. The annuli with the largest radial distance from the galaxy are used to measure extended SBs.}
\label{fig:sb_profiles}
\end{figure*}

To test for the presence of \ion{Mg}{2} emission, we perform aperture photometry on the continuum-subtracted images using the python library \emph{Photutils}. We choose annuli with a radial thickness of 1 pixel or $0.25 ''$, such that the inner radius is $r_{inner}=r_{outer}-1$ (in pixels). Each annulus is centered on the flux-weighted centroid of the galaxy. By dividing the summed flux in each annulus by the area in arcseconds we produce surface brightness (SB) profiles for each galaxy. These profiles are shown in Figure \ref{fig:sb_profiles}. 

The error in the SB is determined from the stacked RMS images of each object.  We adopt annuli that are identical to the annuli used to find the SB profiles for each galaxy. To calculate the variance inside each annulus, we sum the RMS pixel values in quadrature, then divide by the area of each annulus. 

To calculate the $1\sigma$ SB limit we follow the procedure of \cite{Battaia_2015}. We first mask out all the sources, their associated extended halos, and edge noise in both the HeII+47 and HeII/3000+48 images. We then calculate the RMS of the background in randomly-placed $1\arcsec$ apertures. We convert these RMS values to SB limits per $1~\rm arcsec^2$  aperture. We find that the 1$\sigma$ detection limits (SB$_1$) are $6.332\times10^{-19}$ ergs sec $^{-1}$ cm$^{-2}$ arcsec$^2$ and $5.808\times10^{-19} $ ergs sec $^{-1}$ cm$^{-2}$ arcsec$^2$ in the HeII/3000+48 and HeII+47 filters, respectively. 
With the 1$\sigma$ detection limit, SB$_1$, determined for the continuum+\ion{Mg}{2} (HeII+47) image, we define a thicker (or ``extended") annulus to be used to search for any extended \ion{Mg}{2} emission. This annulus will have an inner radius approximately the size of the SB$_1$ isophotal contour for each galaxy. The outer radius is chosen to be the inner radius plus 5 pixels. With this larger annulus, we can average any flux over large areas to reach lower values of SB. The mean radii of these extended apertures are 18, 18, 24 and 14 kpc from the centers of the targets J.26, J.36, J.03 and J.64, respectively. The resulting SB measurements are shown in Figure~\ref{fig:sb_profiles}.

\begin{figure}[!ht]
\centering
\includegraphics[scale=0.6]{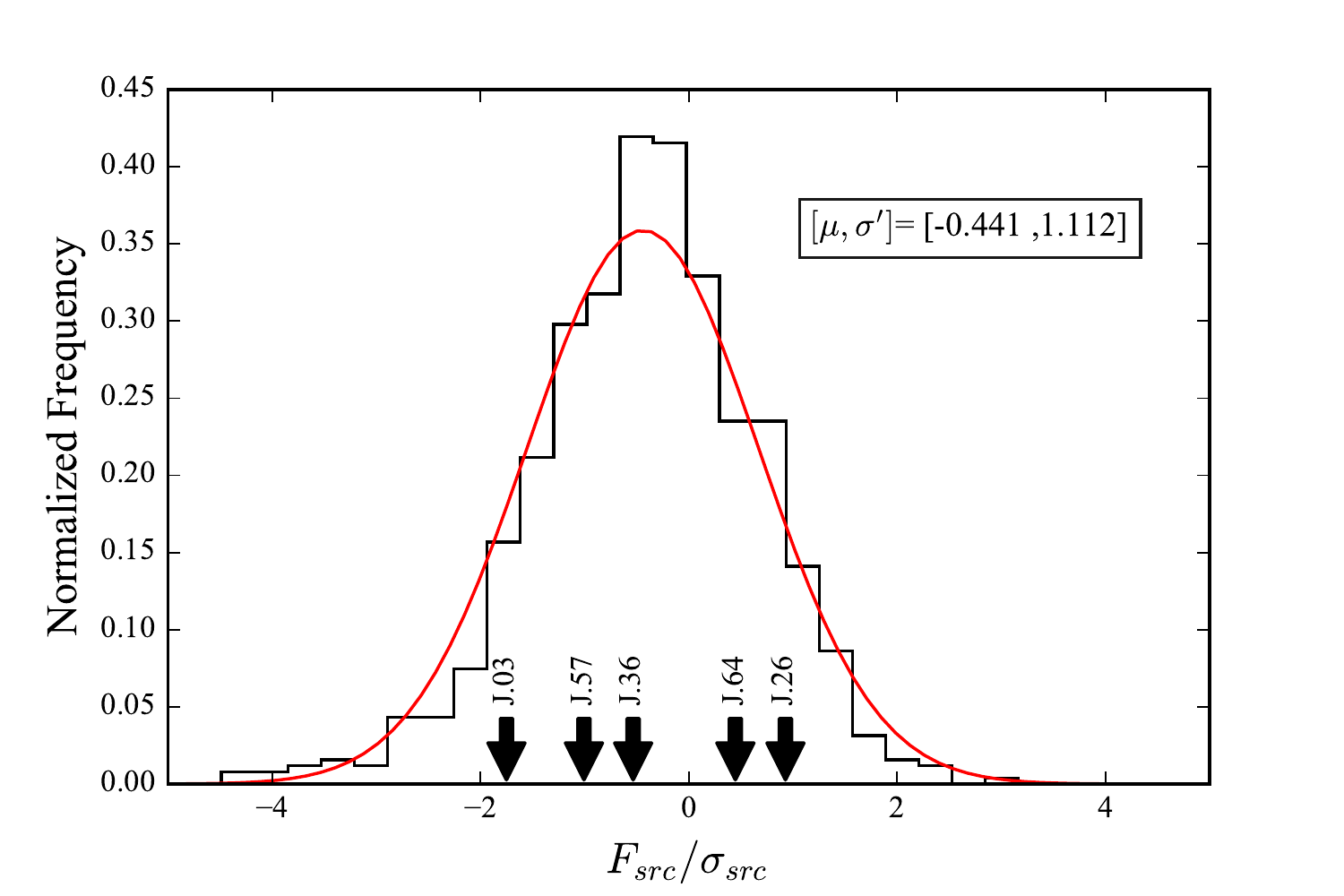}
\caption{Normalized distribution of $F_{\text{src}}/\sigma_{\text{src}}$ values for random circular annuli placed on the continuum-subtracted line image. $F_{\text{src}}$ is the total flux within an aperture and $\sigma_{\text{src}}$ is the expected $1\sigma$ flux limit in the ideal case of perfect sky and continuum subtraction, i.e $SB_{1}\sqrt{A_\text{src}}$. The red line is a Gaussian fit to this normalized distribution with standard deviation, $\sigma=1.112$, and mean, $\mu=-0.441$. The black arrows point to the statistical significance of the flux inside the ``extended annulus'' of each galaxy.}
\label{fig:limits}
\end{figure}

In the case of perfect sky subtraction and continuum subtraction, the 1$\sigma$ SB limit for an extended source is $SB_{1}/\sqrt{A_\text{src}}$, where $A_\text{src}$ is the area in arcsec$^2$ and $SB_{1}$ is the surface brightness limit per 1 arcsec$^2$ aperture. However, our actual detection limits are altered by systematic errors from imperfect subtraction. Therefore, we determine the limits as follows. 
We first mask all the artifacts and sources in the continuum-subtracted images. Next, we generate many apertures with sizes similar to our extended annuli ($\sim 20 \rm\ sq.arcsec$), place them at random, and extract the fluxes, $F_{\text{src}}$, within these apertures. We then normalize the values of $F_{\text{src}}$ by dividing by $\sigma_{\text{src}}$, where $\sigma_{\text{src}} \equiv SB_{1}\sqrt{A_\text{src}}$. For perfect sky subtraction and continuum subtraction, the distribution of extracted fluxes should follow a Gaussian distribution with a standard deviation equal to $\sigma_{\text{src}}$. The distribution of $F_{\text{src}}/\sigma_{\text{src}}$ for these apertures is shown in Figure \ref{fig:limits}

We calculate the standard deviation and mean of the distribution and find that the variance of the distribution is $\sigma'_{\text{src}}=1.1$, implying that the SB detection limit for our continuum-subtracted image is higher than $\sigma_{\text{src}}$ by a factor of 10\%. We adopt $F_{\text{limit}} \equiv \sigma'_{\text{src}}$  as the $1 \sigma$ upper limit on the total line flux of extended \ion{Mg}{2} emission. The SB$_{\text{limit}}$ is then $F_{\text{limit}}/A_{\rm{src}}$. The values of $F_{\text{src}}$ and 5SB$_{\text{limit}}$ for each galaxy are listed in Table \ref{tab:det_lims}.

\begin{deluxetable}{c D c c}
\tablecaption{Significance of extracted flux and detection limits.\label{tab:det_lims}}
\tablehead{\colhead {Object } & \multicolumn2c{F$_{\rm{src}}$(\ion{Mg}{2})\tablenotemark{a}} & \colhead {5SB$_{\text{limit}}$\tablenotemark{b}} & \colhead {Area\tablenotemark{c}}}
\decimals
\startdata
J033225.26-274524.0 &   2.44(0.92)   & 6.51 & 21 \\
J033232.36-274725.0 &  -1.40(-0.53)  & 6.51 & 21 \\
J033230.03-274347.3 &  -5.23(-1.75)  & 5.74 & 27 \\
J033229.64-274242.5 &   1.23 (0.44)  & 6.22 & 26 \\
J033230.57-274518.2 &  -2.53 (-1.00) & 6.81 &18 \\
\enddata
\tablenotetext{a}{\ion{Mg}{2} flux is in units of $10^{-18}$ ergs sec$^{-1}$ cm$^{-2}$. The value in parentheses is the statistical significance with respect to $\sigma_{\text{src}}$.}
\tablenotetext{b}{Limits are in units of $10^{-19}$ ergs sec $^{-1}$ cm$^{-2}$ arcsec$^{-2}$.}
\tablenotetext{c}{Area of the extended annulus in arcsec$^2$.}
\end{deluxetable}

\subsection{Test of Surface Brightness Limits}\label{subsec:test}
To show that \ion{Mg}{2} emission with SB strengths comparable to our limits can be detected in our narrowband imaging, we simulate emission with varying intensities relative to SB$_{\text{limit}}$. For each galaxy, we assign our simulated emission a constant surface brightness corresponding to 1, 3, 5, 10 and 20 times the $1\sigma$ SB$_{\text{limit}}$ inside the largest annulus used (i.e., the extended annulus). We assume Gaussian noise with 1$\sigma$ equal to 1SB$_{\text{limit}}$. Next, we subtract the continuum in the same manner as explained in Section \ref{subsec.cont_sub}. 

To aid in identifying the presence and detectability of extended \ion{Mg}{2} emission we construct a so-called $\chi_{\text{smooth}}$ image for each level of simulated emission following the technique described in \cite{Hennawi2013} and \cite{Battaia_2015}.

To construct the set of smoothed images, we first performed the following operation on the continuum-subtracted images:
\begin{equation}
I_{\text{smooth}}= \text{CONVOLVE[line-continuum]},
\end{equation}
where the CONVOLVE operation indicates convolution of the \ion{Mg}{2} images with a Gaussian kernel with FWHM=1.5 pixels. Next, we computed the sigma image, $\sigma_{\text{smooth}}$, by convolving the propagated error image:
\begin{equation}
\sigma_{\text{smooth}}=\sqrt{\text{CONVOLVE}^2[\sigma^2]},
\end{equation}
where the CONVOLVE$^2$ operation indicates convolution of the image with the square of the Gaussian kernel. The smoothed $\chi$ image, $\chi_{\text{smooth}}$, is then 
the smoothed line image, $I_{\text{smooth}}$, divided by the sigma image, $\sigma_{\text{smooth}}$.

\begin{figure*}[p]
\centering
\includegraphics[scale=1.2]{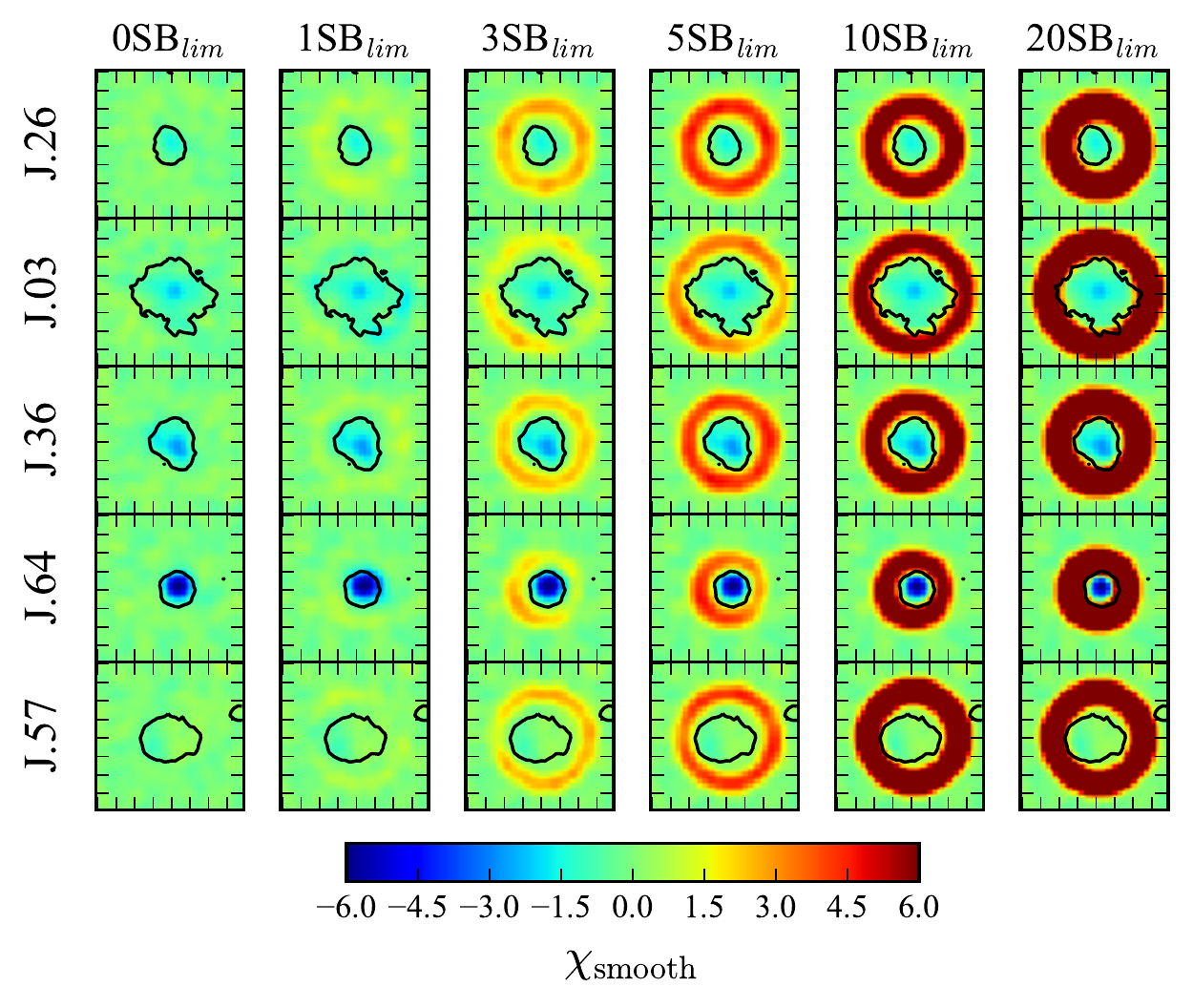}
\caption{Continuum-subtracted $\chi_{\text{smooth}}$ images of the 5 galaxies in our sample. Every galaxy is placed in the same row in each column. The columns show simulated emission, with brightnesses of 0, 1, 3, 5, 10, and 20 times SB$_{\rm lim}$.  Each image has a size of $5'' \times 5''$ (corresponding to $35$ kpc $\times$ $35$ kpc at $z\sim 0.70$). Each image shows the galaxy along with the same isophotal contour used in previous figures (in black).}
\label{fig:sigmas}
\end{figure*}

Figure \ref{fig:sigmas} shows the $\chi_{\text{smooth}}$ images for the 5 levels of simulated \ion{Mg}{2} emission. We also include the $\chi_{\text{smooth}}$ image of each galaxy without any simulated emission (in the left-most column). The galaxies are outlined by a black isophotal contour corresponding to 1SB$_1$ and the simulated emission is contained inside the extended annulus surrounding each contour. The  $\chi_{\text{smooth}}$ images confirm that we should be able to detect extended \ion{Mg}{2} emission down to a conservative level of 5SB$_{\text{limit}}$. 


\subsection{Equivalent Widths}\label{subsec.ew}
Here we derive an expression to calculate the equivalent width (EW$_{\rm{MgII}}$) of any absorption or emission features observed in our narrow-band imaging. Starting from the expression for EW used in the context of spectroscopy,
\begin{equation}
EW_{\lambda}=\int (1-\frac{f_{\lambda}}{f_{\rm{cont}}})d\lambda
\label{eq:specEW}
\end{equation}
we begin by dividing Eq \ref{eq:subtraction} by the flux density of the continuum and the effective width of the on-line filter,
\begin{equation}
\frac{F_{\rm{line}}}{f_{\rm{cont}}\Delta \lambda_{\rm{MgII}}}=\frac{F_{\rm{MgII}}}{f_{\rm{cont}}\Delta \lambda_{\rm{MgII}}}- 1.
\end{equation}
Next, we rearrange the above expression such that we produce the argument of the integrand in Eq. \ref{eq:specEW} on the right hand side,
\begin{equation}
-\frac{F_{\rm{line}}}{f_{\rm{cont}}\Delta \lambda_{\rm{MgII}}}=1-\frac{f_{\rm{MgII}}}{f_{\rm{cont}}}.
\end{equation}
We then approximate the integration in Eq. \ref{eq:specEW} by multiplying the integrand above by the effective width of the on-line filter $d\lambda=\Delta \lambda_{\rm{MgII}},$
\begin{equation}
-\frac{F_{\rm{line}}}{f_{\rm{cont}}}=(1-\frac{f_{\rm{MgII}}}{f_{\rm{cont}}})\Delta \lambda_{\rm{MgII}};
\end{equation}
such that
\begin{equation}
{\rm EW}_{\rm{MgII}}=-\frac{F_{\rm{line}}}{f_{\rm{cont}}}.
\end{equation}

\begin{figure*}
\centering
\gridline{\fig{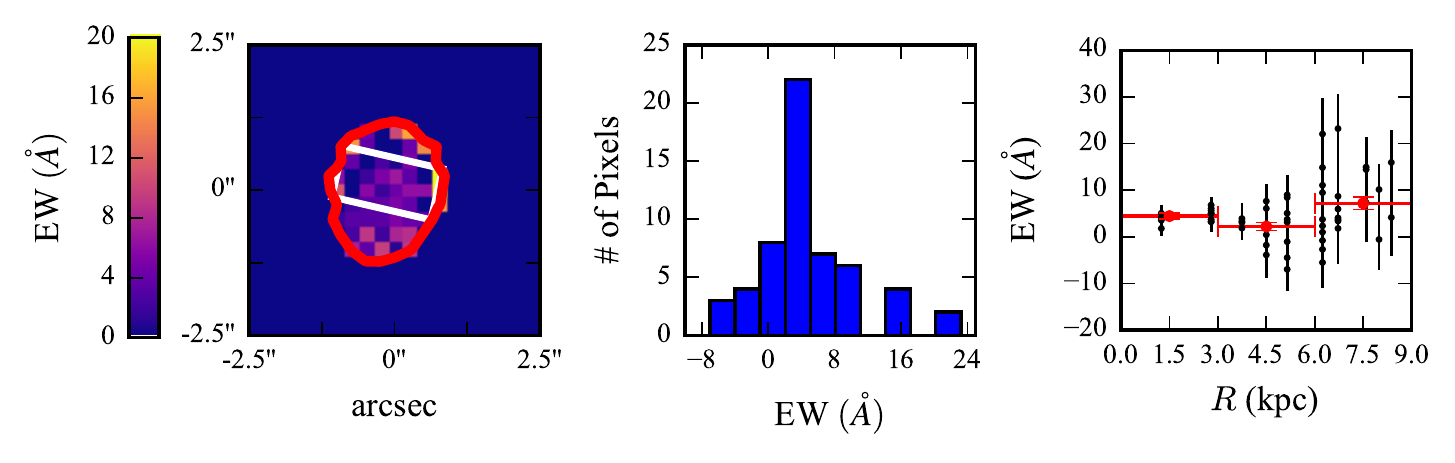}{0.8\textwidth}{(a: J.26)}}
 \gridline{\fig{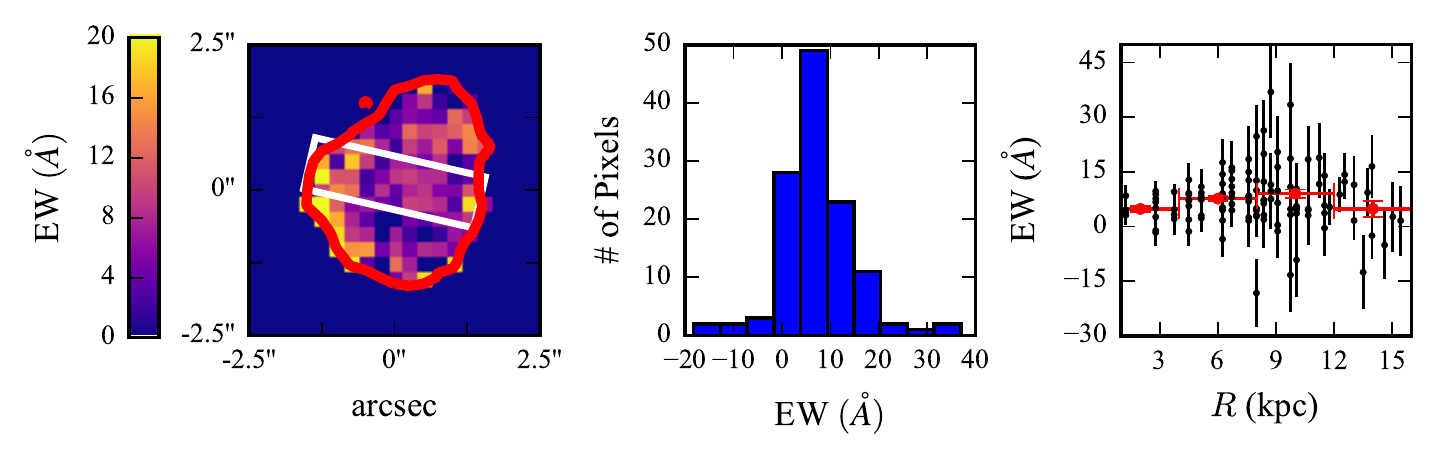}{0.8\textwidth}{(b: J.36)}}
\gridline{\fig{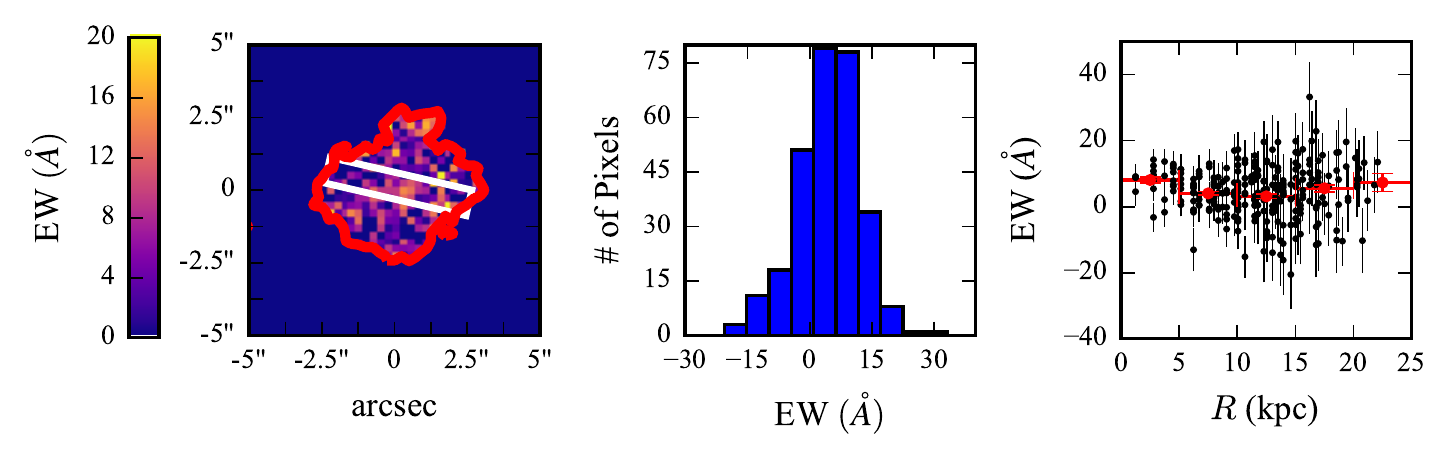}{0.8\textwidth}{(c: J.03)}}
 \gridline{\fig{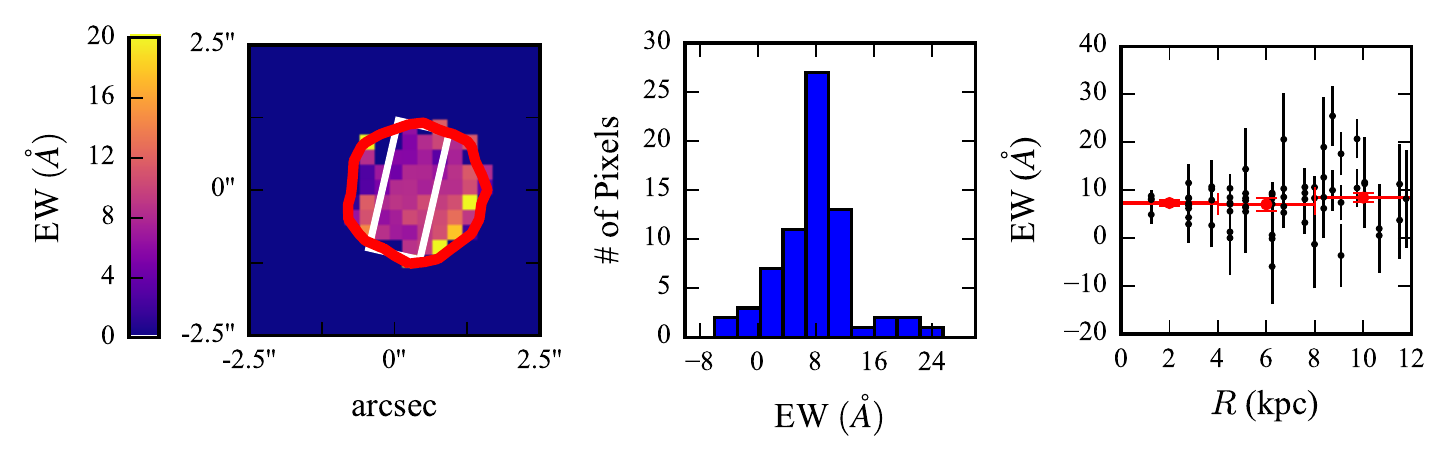}{0.8\textwidth}{(d: J.64)}}
\caption{Left: Images of the equivalent widths EW$_{\text{\ion{Mg}{2}}}$ inside the 1SB$_1$ contour for each galaxy (red). The white contour shows the placement and orientation of the 0.9\arcsec-wide slit used to measure the equivalent width of absorption in the Keck/LRIS spectrum of each galaxy. Middle: Distribution of  EW$_{\text{\ion{Mg}{2}}}$ values in pixels with continuum S/N $>1.5$ inside the slit aperture. Right: The EW$_{\text{\ion{Mg}{2}}}$ values of these high-S/N pixels vs. projected distance from the galaxy center are shown in black, and binned EW$_{\text{\ion{Mg}{2}}}$ measurements are shown in red. The horizontal error bars represent the width of the radial bin used.}
\label{fig:ews}
\end{figure*}

Using the above equation along with the continuum and continuum-subtracted images, we produce images of the observed-frame EW$_{\rm{MgII}}$. 
They are displayed for each galaxy in Figure \ref{fig:ews} and show only the EWs within the 1$\sigma$ SB$_1$ contours of the corresponding \ion{Mg}{2} images (prior to continuum subtraction). 

To compare our map of EW$_{\text{\ion{Mg}{2}}}$ to the values measured from the Keck/LRIS spectra, we place 0.9 arcsec-wide slit-like apertures over each galaxy. The width and position angle of the apertures are consistent with the orientation of the slits used to obtain the spectra. Next, we determine which pixels lie outside the 1$\sigma$ SB$_1$ contours and set their values to zero. Outside this contour, the EW$_{\text{\ion{Mg}{2}}}$ values become poorly constrained due to the lack of S/N in the continuum. We then select all pixels with a S/N $\ge 1.5$ within 
each slit aperture
and create a histogram to show the distribution of their EW$_{\text{\ion{Mg}{2}}}$ values. The histograms are shown in Figure \ref{fig:ews}. We also compute the mean equivalent width of these pixels and report their values in Table \ref{tab:abs_props}.
 
To assess the morphology of the \ion{Mg}{2} absorption, we determine the projected distance of each pixel from the center of each galaxy in kiloparsecs. We plot the EW$_{\rm{MgII}}$ vs. this projected distance for each galaxy in the right panels of Figure \ref{fig:ews}. Although some of the plots suggest a slight upward trend in the values of EW$_{\rm{MgII}}$ with increasing radii, we cannot be confident in this trend because of the large scatter. To better visualize the data and test the significance of the trend, we bin the data radially in bins with widths between 3 and 5 kpc. For example, in Figure \ref{fig:ews}(c), 
EW$_{\text{\ion{Mg}{2}}}$ values shown for J.03 extend 
out to $\sim$ 25 kpc (shown in black in the right-most panel). We bin these EW$_{\rm{MgII}}$ values in 5 kpc increments. We calculate the mean and scatter of the EW$_{\text{\ion{Mg}{2}}}$ in each bin and show these values in red in the right panels of Figure \ref{fig:ews} and in Figure \ref{fig:ew_comb}.   

\section{Results}\label{sec:results}

\subsection{Limits on \ion{Mg}{2} Emission}
For our sample of galaxies, we are sensitive to emission in our ``extended'' annuli with mean distances of 18, 18, 24 and 14 kpc from the centers of J.26, J.36, J.03 and J.64, respectively.
 We do not detect any significant \ion{Mg}{2} emission at these distances around any of our target galaxies. The $\chi_{\text{smooth}}$ images shown in Figure \ref{fig:sigmas} confirm this. A comparison of the simulated emission with the $\chi_{\text{smooth}}$ version of the original continuum-subtracted image, shown in the first column, similarly suggests that we do not detect any extended \ion{Mg}{2} emission. We thus place $5\sigma$ upper limits on the SB of \ion{Mg}{2} emission for each galaxy in the sample, summarized in Table \ref{tab:det_lims}. The most sensitive detection limit using the largest area is SB(\ion{Mg}{2}) $<$ 5.74 $\times$ $10^{-19}$ ergs sec $^{-1}$ cm$^{-2}$ arcsec$^{-2}$, computed for the galaxy J.03. 

\subsection{Spatially Resolved Maps of \ion{Mg}{2} Absorption}
In this section we discuss the details of the absorption detected in our SB profiles as well as compare our EW$_{\rm{MgII}}$ measurements to those measured in the supplemental Keck/LRIS spectra. 

\subsubsection{Effects of \ion{Mg}{2}\ Absorption on Surface Brightness Profiles} \label{subsubsec:SBprofiles}
Although we do not detect any extended \ion{Mg}{2} emission, we do observe a decrement of flux, SB$_{\rm abs}$, in the SB profiles of 4 out of 5 galaxies in our sample. As shown in Figure \ref{fig:sb_profiles}, absorption from \ion{Mg}{2} ions is prevalent in the profiles at projected distances $\lesssim5$ kpc, and decreases radially outward from the maximum absorption at the center of the galaxies. In Table \ref{tab:abs_props} we report for the galaxies J.26, J.36 and J.03 a maximum decrement in the SB profile due to absorption (SB$_{\rm abs}$) $\approx$ -5 $\times10^{-18}$ erg s$^{-1}$ cm$^{-2}$ arcsec$^{-2}$. Additionally, we report for galaxy J.64 a SB$_{\rm abs}$ with a significantly more negative value of $(-18.2 \pm 0.1) \times10^{-18}$ erg s$^{-1}$ cm$^{-2}$ arcsec$^{-2}$. Finally, for J.57, 
the value of SB$_{\rm abs} = (-1.25 \pm 1.12) \times10^{-18}$ erg s$^{-1}$ cm$^{-2}$ arcsec$^{-2}$, and is consistent with measuring zero absorption as expected given the redshift of this system. This measurement suggests that the quality of our continuum subtraction is satisfactory.

\subsubsection{Morphology of MgII Absorption}
Figure \ref{fig:ews}  shows the images, distributions and radial projections of \ion{Mg}{2} EWs. We have zeroed out any values that lie outside the SB$_1$ contours for each galaxy. We also impose a signal-to-noise cut, only including EW$_{\text{\ion{Mg}{2}}}$ values in the middle and right panels for pixels in which the continuum S/N is greater than $1.5$ and which are inside each Keck/LRIS aperture, defined in Sec. \ref{subsec.ew}. 
The mean EW$_{\text{\ion{Mg}{2}}}$ is computed for all pixels inside these apertures 
and the error is propagated in quadrature. The resulting values of the mean EW and the error in these measurements are summarized in Table \ref{tab:abs_props}. Comparing our narrowband EWs with those measured from the spectra, we find agreement to within  1.6-4.6$\sigma$ for galaxies J.26 and J.36, and more statistically significant differences for galaxies J.03 and J.64. We discuss possible causes for these differences below.  

Given the size of the median seeing disk for these observations ($\rm FWHM\approx0.8\arcsec$), the EW$_{\text{\ion{Mg}{2}}}$ values measured in adjacent pixels (each of which subtends $0.25\arcsec$) are not independent, and hence their errors are covariant. This covariance implies that the uncertainties in our mean EW$_{\text{\ion{Mg}{2}}}$ values are underestimated, such that the discrepancies between these values and those measured in our LRIS spectra are likely less significant than the tension described above. 

Figure \ref{fig:spec_images}(c) shows the Keck/LRIS spectrum of galaxy J.03. The continuum observed near the \ion{Mg}{2} transition has low S/N compared to the spectra of the rest of the sample. Since the value of EW$_{\text{\ion{Mg}{2}}}$ depends on the level of the continuum, 
it may well be that our choice of continuum level in calculating the EW$_{\text{\ion{Mg}{2}}}$ from the spectrum is higher than the continuum level implied by our narrow-band image.  Such a systematic error could give rise to a higher spectroscopic EW.

Figure \ref{fig:spec_images}(d) shows the Keck/LRIS spectrum of galaxy J.64. This object is the brightest galaxy in the sample, and also exhibits the highest-velocity wind.  This shifts the \ion{Mg}{2} absorption profile toward the blue end of the 
HeII+47 transmission curve, which could cause the signal in this filter to be dominated by the continuum level and the absorption signal to be underestimated.

\begin{figure*}[!htb]
\centering
\includegraphics[scale=0.7]{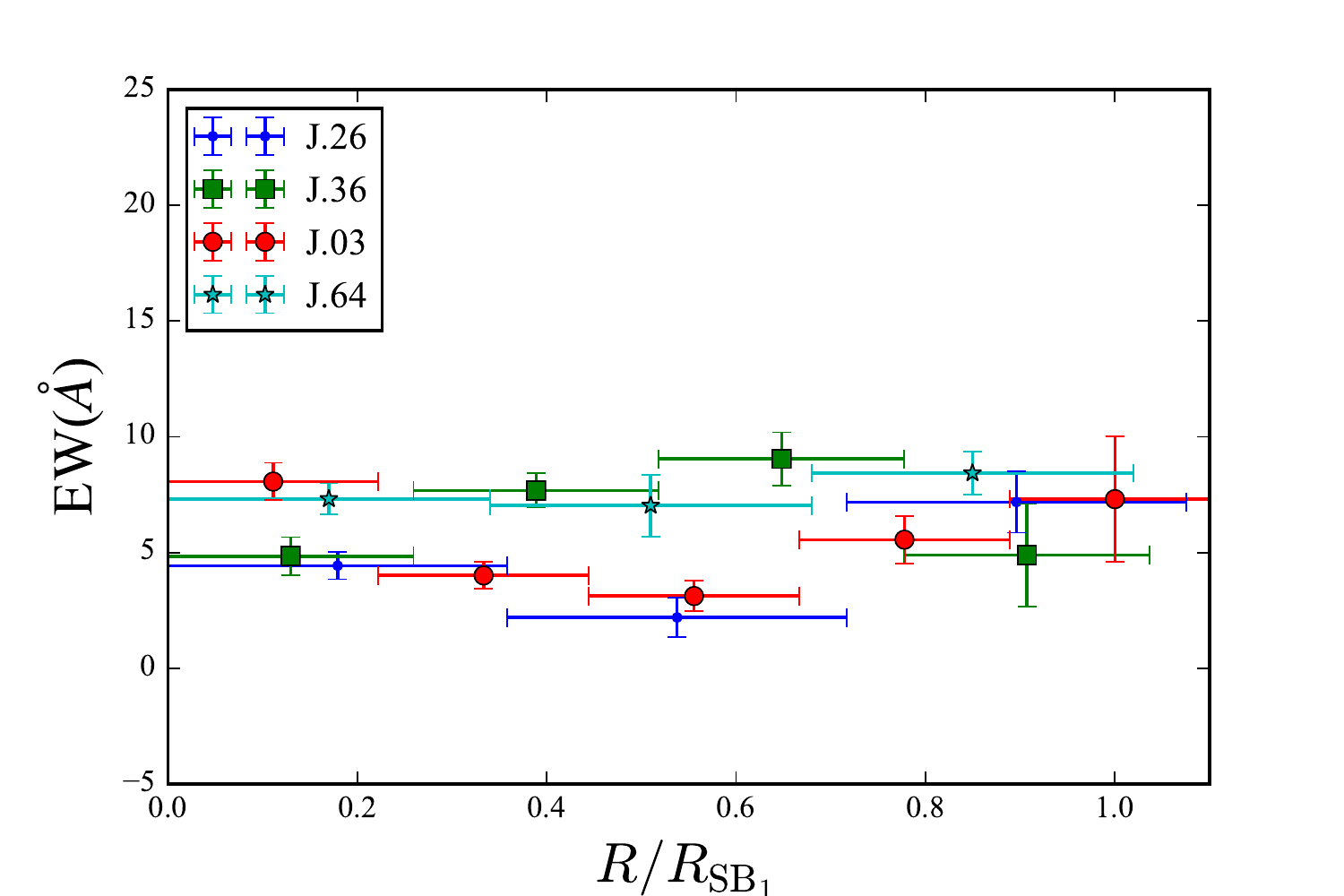}
\caption{Radial profile of the mean EW of \ion{Mg}{2} absorption for four of our sample galaxies. Profiles for J.26,  J.36,  J.03, and J.64 are shown with small blue circles, green squares, large red circles, and cyan stars, respectively. The mean EW values are the same as those shown in the right column of Figure \ref{fig:ews}. We have normalized the corresponding radii by the approximate size of the SB$_1$ contour of each galaxy.}
\label{fig:ew_comb}
\end{figure*}

As demonstrated in 
the right panels of Figure \ref{fig:ews}, 
a majority of the galaxies exhibit large scatter in 
EW$_{\text{\ion{Mg}{2}}}$ at large radii. To better understand the significance of any possible trends in these values, 
we compile the mean EW$_{\text{\ion{Mg}{2}}}$ values for all the galaxies and show their profiles in Figure \ref{fig:ew_comb}. To account for the varying sizes of the galaxies, we normalize the radii of the bins by the approximate radius of the SB$_1$ contour for each galaxy. 
Upon inspection of this figure, we see that the galaxies exhibit no statistically significant trend in the mean absorption EW$_{\rm MgII}$ as a function of radius inside our 1SB$_1$ isophotal contour, which suggests that the covering fraction 
of saturated \ion{Mg}{2} absorption
is approximately constant across the surface.

\begin{deluxetable}{c c c c c}
\tablecaption{Properties of \ion{Mg}{2}\ Absorption\label{tab:abs_props}}
\tablehead{\colhead {Object } & \colhead {SB$_{\rm abs}$\tablenotemark{a}} & \colhead {$R_{\perp}^{\rm{SB}_1}$\tablenotemark{b}} & \colhead {LRIS EW$_{\text{\ion{Mg}{2}}}^{\rm obs}$\tablenotemark{c}} & \colhead {NB EW$_{\text{\ion{Mg}{2}}}^{\rm obs}$\tablenotemark{d}} \\
\colhead{} & \colhead{} & \colhead{(kpc)} & \colhead{(\AA)} & \colhead{(\AA)}}
\startdata
J.26 & \ $-5.4 \pm 1.2 $  & 8   &\  $7.5 \pm 0.4$ & \ \ $3.5 \pm 0.8$ \\
J.36 & \ $-5.6 \pm 0.1 $  & 15 &\  $5.8 \pm 0.5$ & \ \ $7.7 \pm 1.0$ \\
J.03 &\  $-4.4 \pm 0.6 $  & 21 & $12.7 \pm 1.7$ & \ \ $5.4 \pm 0.7$  \\
J.64 & $-18.2 \pm 0.1$ & 10 & $13.2 \pm 0.3$   &\ \ $7.6 \pm 0.7$ \\
J.57 &\  $-1.3 \pm 1.2 $  & 11 & $6.10 \pm 0.4$ & $-0.7 \pm 0.6$ \\
\enddata
\tablenotetext{a}{Maximum SB decrement in units of $10^{-18}$ erg sec $^{-1}$ cm$^{-2}$ arcsec$^{-2}$.}
\tablenotetext{b}{Radius of SB$_1$ contour.}
\tablenotetext{c}{Measured from Keck/LRIS spectra in the observed frame. Includes both lines in the \ion{Mg}{2} doublet.}
\tablenotetext{d}{Measured from narrowband images, and reported in the observed frame.}
\end{deluxetable}

\section{Discussion}\label{sec:discussion}
\subsection{Previous Detections of Extended \ion{Mg}{2} Emission}
Previous constraints on the brightness of scattered \ion{Mg}{2} emission were reported by \cite{Rubin_2011} and \cite{Martin2013}. In \cite{Rubin_2011} the authors studied emission from the starburst galaxy TKRS 4389 at $z = 0.69$ with a SFR of $49.8\msunperyr$. This emission was detected in a 2-dimensional Keck/LRIS spectrum, with flux from the emission reaching $(8.0 \pm 0.4)$ and $(4.4 \pm 0.4)$ $\times10^{-18}$ ergs sec$^{-1}$ cm$^{-2}$ at  $\lambda _{2796}$ and $(4.0 \pm 0.3)$ and $(2.5 \pm 0.4)$ $\times10^{-18}$ ergs sec$^{-1}$ cm$^{-2}$ at $\lambda_{2803}$ in two independent locations spatially offset from the galaxy continuum. The flux from both emission lines can be converted into two surface brightness values by taking the average of the flux measured at each location and each transition, and dividing by a 1 arcsec$^2$ aperture. 

A second detection of extended \ion{Mg}{2} emission is reported in \cite{Martin2013}. In this study, the authors spatially resolve extended \ion{Mg}{2} emission in the galaxy 32016857 at a redshift of $z=0.9392$ with a SFR of $\sim 80\msunperyr$. Detected using a 2-dimensional Keck/LRIS spectrum, the \ion{Mg}{2} emission extends out to $\sim11$ kpc, or 1.4\arcsec\ at $z\sim 0.94$, away from the galaxy continuum to the East. In the integrated spectrum, the observed flux in \ion{Mg}{2} emission is approximately $1.5$$\times10^{-17}$ ergs sec$^{-1}$ cm$^{-2}$ at  $\lambda _{2796}$ and $1.0$$\times10^{-17}$ ergs sec$^{-1}$ cm$^{-2}$ at $\lambda_{2803}$ to 20\% accuracy. The extended \ion{Mg}{2} component contributes up to 46\% of the total integrated \ion{Mg}{2} flux.  As for TKRS4389, we calculate a SB  assuming that the total extended \ion{Mg}{2} emission flux is $0.46\times2.5$$\times10^{-17}$ ergs sec$^{-1}$ cm$^{-2}$ and that it covers an area of 1.4\arcsec\ times the 1.2\arcsec\ Keck/LRIS slit width. Figure \ref{fig:all_limits} shows 5$\sigma$ SB detection limits for each galaxy in our sample and the SB calculated for the galaxies TKRS 4389 and 32016857 vs. SFR (left panel) and vs.\ $\log M_*/M_{\odot}$ (middle panel). These figures suggest that we should be able to detect scattered \ion{Mg}{2} emission with strengths similar to that detected in both TKRS 4389 and 32016857 in our narrowband imaging.

\subsection{Possible Correlation of \ion{Mg}{2} Emission Strength with Galaxy Properties }
\begin{figure*}[!htb]
\centering
\includegraphics[scale=0.57]{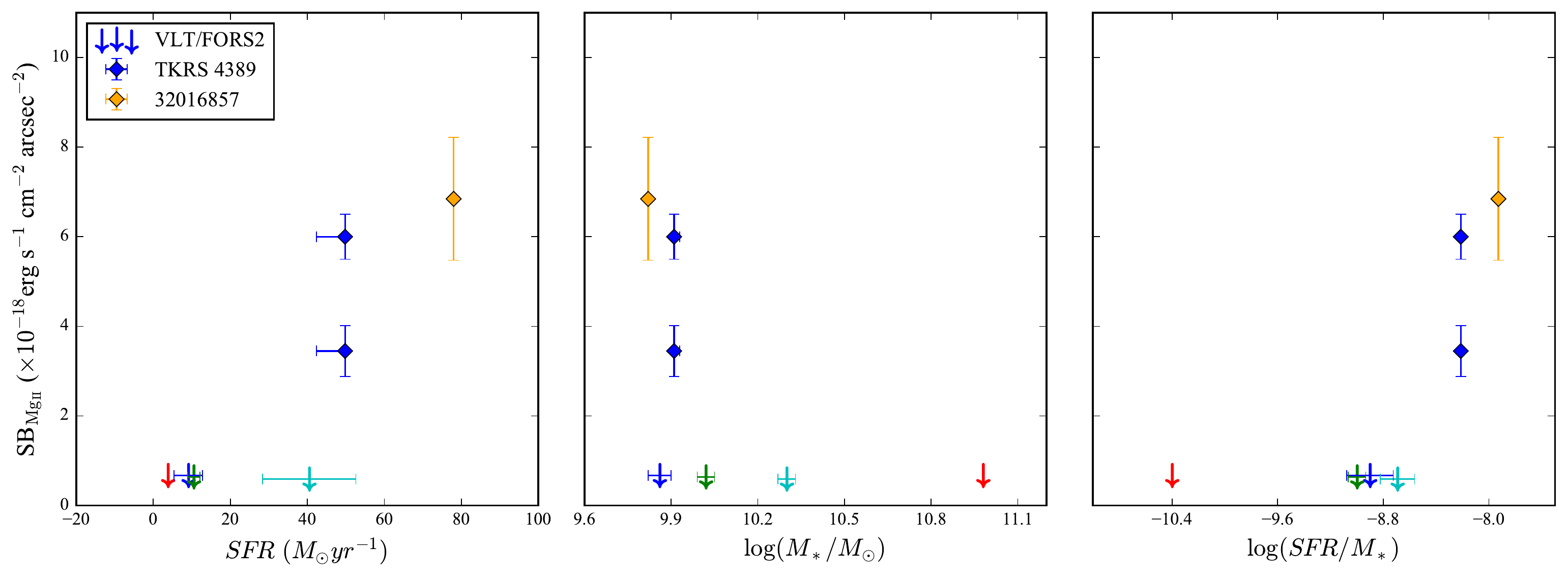}
\caption{Comparison of our detection limits to the SB of extended \ion{Mg}{2} emission measured in two independent locations spatially offset from the continuum around the starburst galaxy TKRS 4389 (reported by \citealt{Rubin_2011}), and in one location spatially offset from the continuum eastward of the star-forming galaxy 32016857 (reported by \citealt{Martin2013}). These surface brightness constraints are plotted vs.\ SFR (left), stellar mass (middle), and specific SFR (right). Our imaging is sufficiently sensitive to detect extended emission at similar strengths to the extended emission measured for TKRS 4389 with SFR $\sim 50 \msunperyr$ and 32016857 with SFR $\sim 80 \msunperyr$. The symbol colors indicate the same galaxies as in Figure \ref{fig:ew_comb}.}
\label{fig:all_limits}
\end{figure*}

Taken at face value, the left panel of Figure \ref{fig:all_limits} could be consistent with a positive correlation between \ion{Mg}{2} SB and SFR. 
The constraints shown in the middle panel of Figure \ref{fig:all_limits} are similarly suggestive of (and consistent with) a negative correlation between the SB of extended \ion{Mg}{2} emission and galaxy stellar mass. 
Future observations are needed to verify these trends, and the possibility that objects with yet higher SFRs ($\gtrsim50 \msunperyr$) exhibit brighter extended \ion{Mg}{2} emission.  
Finally, the right panel of Figure \ref{fig:all_limits} shows that the galaxy with detected extended \ion{Mg}{2} also has the highest specific SFR. If it is ultimately confirmed that low-$M_*$ galaxies with the highest SFRs, or highest specific SFRs, exhibit the brightest emission, this could point to a physical link between the escape velocity of galaxies and the spatial extent and/or optical depth of wind material.
Here we note that a similar trend was observed by both \cite{Erb2012} and \cite{Feltre2018} in their examination of the total \ion{Mg}{2} emission strength vs. $M_*$. 

A potential complicating factor in the interpretation of trends in \ion{Mg}{2} SB with galaxy properties is the possible contribution of nebular emission to the \ion{Mg}{2} line profiles.
Recent studies by \cite{Henry2018} and \cite{Guseva2019} suggest that there may be a significant contribution from nebular emission (i.e., from \ion{H}{2} regions) to the total \ion{Mg}{2} emission strength in some galaxies. 
In detail, \cite{Henry2018} report strong \ion{Mg}{2} emission having 
$\rm EW_{2796}\sim 0.4-9.1$ \AA\ along with negligible absorption
arising in a sample of extreme compact starburst galaxies (``Green Peas'') at $z \sim 0.2-0.3$. 
Their photoionization modeling of the emission line strengths suggests that the observed 
\ion{Mg}{2} emission fluxes can in fact be 
dominated by \ion{H}{2} region emission in such low-metallicity systems (with $Z = 0.16-0.32 Z_{\odot}$).  They also note that higher metallicity and/or more dusty conditions will produce weaker nebular emission.  Moreover, these authors find that the strength of observed \ion{Mg}{2} emission is correlated with the escape fraction of \ion{Mg}{2} photons, suggesting that within their sample, galaxies with the weakest observed emission may have the strongest intrinsic nebular emission. 

Therefore, our galaxies -- exhibiting weaker emission in comparison to those studied by  \citet{Henry2018} -- might have \ion{H}{2} regions producing significant intrinsic \ion{Mg}{2} emission.
However, our sample (along with TKRS4389) is also $\gtrsim1$ dex higher in stellar mass and thus richer in both metals (by $\sim +0.9$ dex in $\log$ O/H; \citethnop{Zahid2011}) and dust than the \citet{Henry2018} Green Peas.
While there may indeed be a nebular contribution to \ion{Mg}{2} emission when it is observed in galaxies in this mass range,  
more detailed photoionization modeling is required to estimate line luminosities for the relevant physical conditions.  As nebular emission is reprocessed by scattering just as continuum photons are, its presence would tend to brighten any spatially-extended line component, and any trends in nebular emission line strengths with $M_*$ or SFR would be reflected in the surface brightnesses of extended emission.
The nebular contribution should therefore be considered by future studies interpreting the meaning of putative trends in extended line SB with galaxy properties.

\subsection{Geometry of Scattering Material}
In the context of the idealized models of cool gas outflows discussed in \cite{Prochaska_2011}, radiative transfer calculations predict that strong \ion{Mg}{2} emission will always accompany the blueshifted \ion{Mg}{2} absorption that is ubiquitously observed to trace galactic-scale winds \citep{Weiner2009,Martin2012,Rubin_2014}. For isotropic and dust-free scenarios, photons are conserved, as any absorbed continuum photon is eventually re-emitted. Therefore, the total equivalent width of both the absorption and emission features is equal to zero in such models. Assuming that our galaxies host an isotropic and dust-free wind (and that they do not produce significant nebular \ion{Mg}{2} emission), we wish to determine how much emission is predicted to be generated by this wind, and how the SB of this emission compares to our detection limits.

To calculate the predicted emission flux we first determine the flux absorbed by \ion{Mg}{2} ions. Using our Keck/LRIS spectra, we find the average value of the continuum near the \ion{Mg}{2} doublet and multiply this value by the observed EW of the doublet. Then to estimate the SB, we distribute this flux uniformly inside multiple annuli of varying sizes. These annuli all have an inner radius equal to the galaxy's isophotal radius and successively larger outer radii.  Additionally, since our SB limits are dependent on the size of the aperture used, we calculate the SB detection limits of our images inside each of the aforementioned annuli. Figure \ref{fig.emission} shows how the predicted SB of emission varies with the spatial extent of the annulus (red octagons), as well as how the SB compares with our detection limits (thin black curve). Excepting galaxy J.03, the predicted SB of this emission lies above our detection limits. 
Under the assumption that the wind in these galaxies does in fact extend beyond the $\rm SB_1$ isophotal contour (at $R_{\perp}^{\rm{SB}_1} = 8-21$ kpc), the absence of the predicted emission 
in our narrowband imaging suggests that these galaxies do not host isotropic, dust-free winds.  

\subsubsection{Anisotropic, Dust-Free Winds}
There are many phenomena that may reduce the SB of the scattered \ion{Mg}{2} emission so that it is consistent with our observations. One factor that can affect the observed emission strength is the morphology of the wind. Anisotropic winds were shown in \cite{Prochaska_2011} to exhibit reduced emission strengths compared to isotropic winds. Direct evidence for anisotropic winds, and specifically for a bipolar morphology, has been observed in emission from cold and shock-heated gas around local starburst galaxies \citep[e.g.,][]{{Walter2002,Westmoquette2008M,Strickland2009}}. Around distant galaxies, enhanced \ion{Mg}{2} absorption along a galaxy's minor axis \citep[][]{{Bordoloi2011,Kacprzak2012,Bouche2012}} observed toward background QSO sightlines is likewise suggestive of bipolar outflows. Furthermore, the analysis of \cite{Rubin_2014} demonstrating a strong dependence of the incidence of winds observed ``down the barrel'' on galaxy orientation 
was interpreted as additional, strong evidence for such a morphology.

We now assume that the brightness of emission in our galaxies is reduced by the effect of anisotropy. For the anisotropic winds modeled in \cite{Prochaska_2011}, the emission is reduced by the factor $\Omega/4\pi$, where $\Omega$ is the angular extent of the wind. As \citet{Prochaska_2011} noted, given that the outflow must cover most of the continuum in order to be detected in typical down-the-barrel spectroscopy, the value of $\Omega$ has an approximate lower limit of $\Omega > 2\pi$. We show the predicted SB profiles for wind emission from our galaxies assuming  $\Omega = 2\pi$ with gray diamonds in Figure  \ref{fig.emission}.

After reducing the SB of the expected \ion{Mg}{2} emission by the corresponding factor of 2, we predict profiles that fall below our SB detection limits for galaxies J.26 and J.36. However, the SB profile of J.64 remains above our detection limits, suggesting additional phenomena are needed to reduce the strength of scattered emission. As discussed in Section \ref{sec:results}, this object is the brightest in our sample and exhibits the strongest \ion{Mg}{2} absorption, which suggests the presence of a strong ISM component. \cite{Prochaska_2011} noted that \ion{Mg}{2} photons can be more effectively trapped in such objects with large amounts of dusty interstellar material. 

\subsubsection{Anisotropic, Dusty Winds}

Dust in the wind is another factor that can reduce the observed emission strength and affect the shape of the \ion{Mg}{2} line profile. 
In the \cite{Prochaska_2011} models that include dust in the wind material, the dominant effect is that the most redshifted emission is suppressed. The line flux is reduced by a factor of $(1+\tau_{\rm{dust}})^{-1}$, where $\tau_{\rm{dust}}$ is the integrated opacity of dust. 

\begin{figure*}[h]
\centering
\gridline{\fig{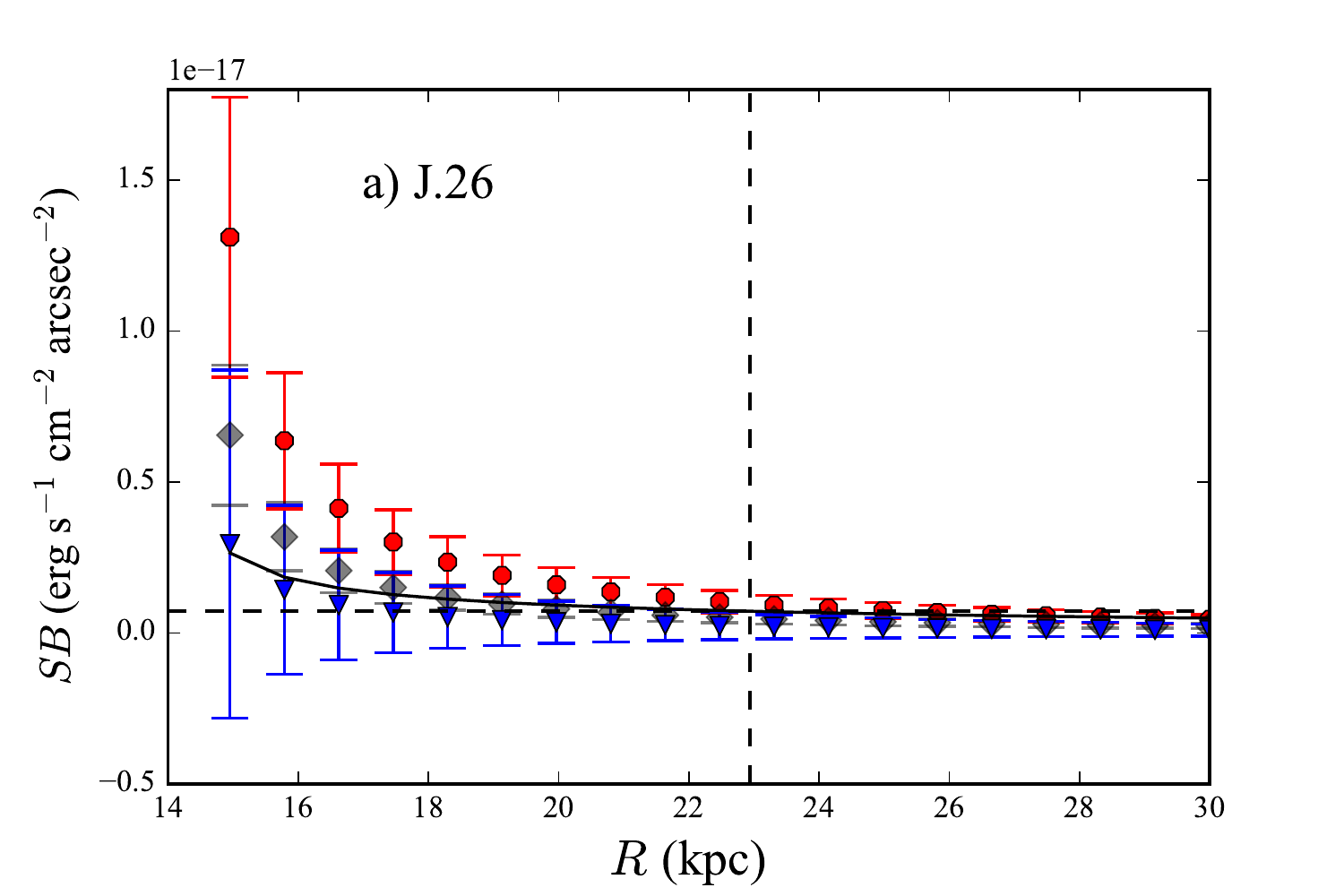}{0.5\textwidth}{(a)}
          \fig{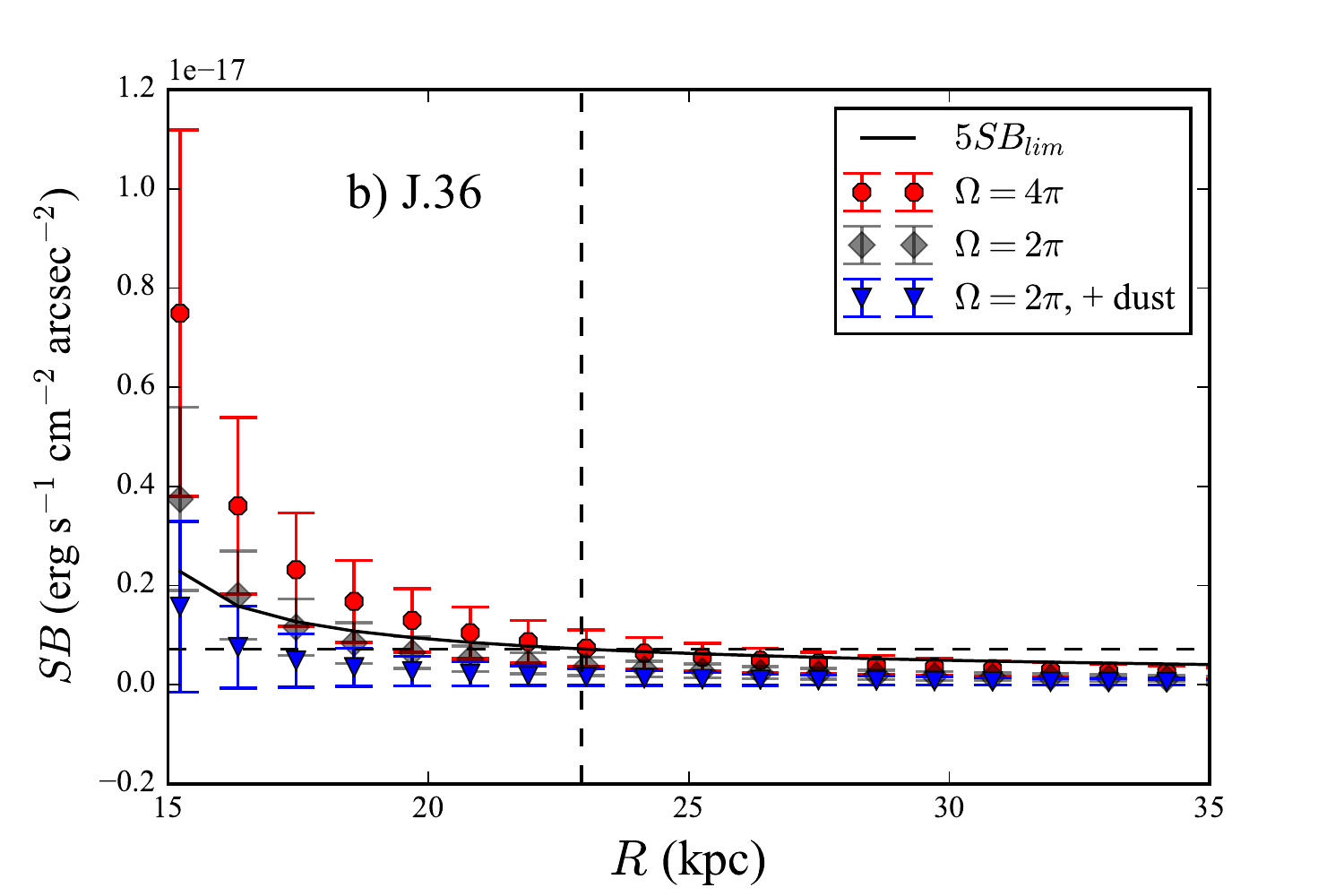}{0.5\textwidth}{(b)}}
\gridline{\fig{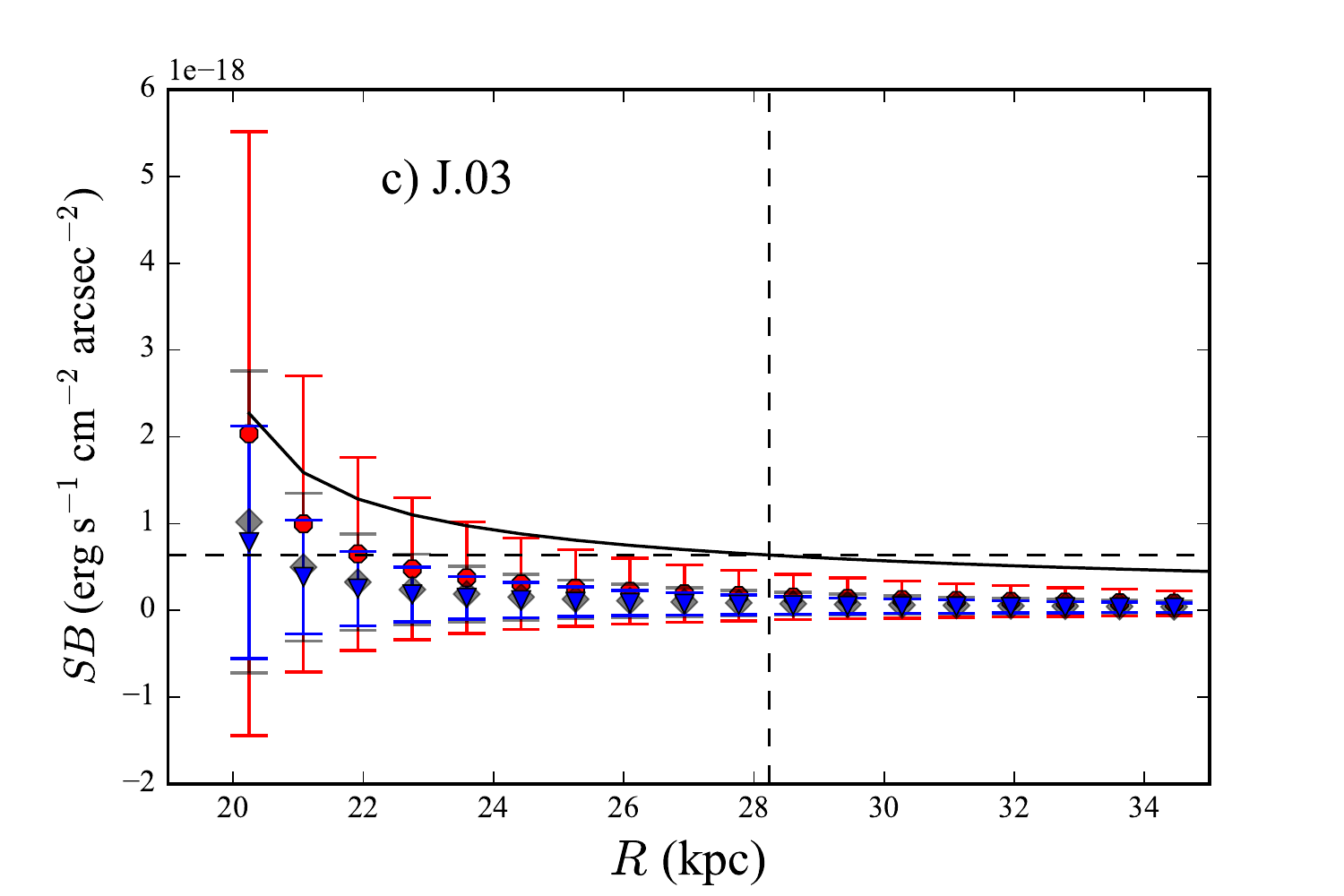}{0.5\textwidth}{(c)}
          \fig{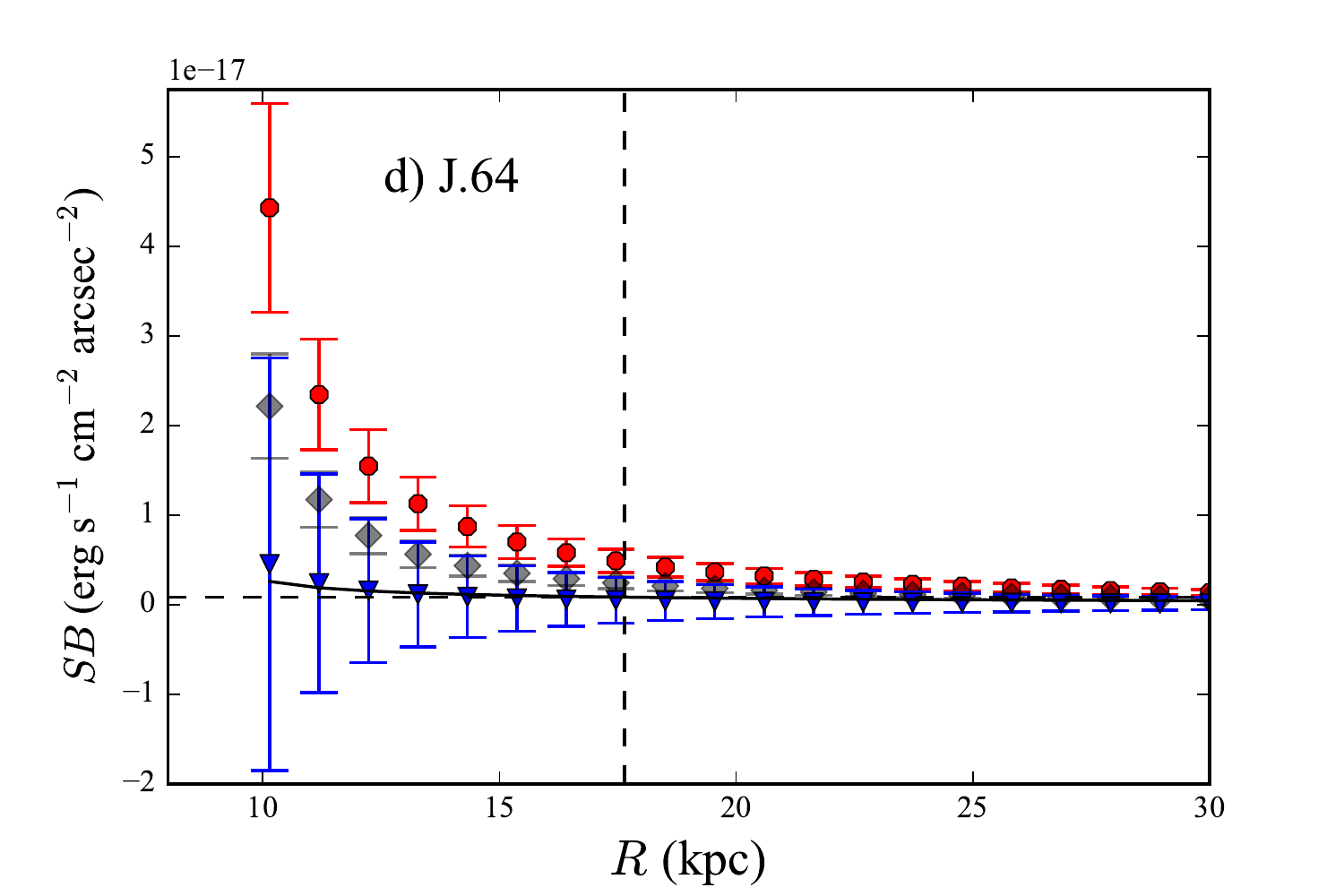}{0.5\textwidth}{(d)}}
\caption{Predicted SB of \ion{Mg}{2} emission for isotropic, anisotropic, and both anisotropic and dusty winds. The red octagons show the expected SB of emission that has been uniformly distributed inside an annulus (with inner radius equal to that of the galaxy's SB$_1$ contour and an outer radius equal to the x-axis value) for an isotropic wind. The gray diamond points show the SB of \ion{Mg}{2} emission predicted for an anisotropic wind with angular extent $\Omega=2\pi$. The blue triangles mark the predicted SB for a wind that is both anisotropic and affected by dust as described in the text. The solid line shows the value of our SB detection limits. The dashed vertical and horizontal lines represent the outer radius of the extended annulus used to measure the primary detection limits reported for each galaxy and the value of that limit. The legend in panel (b) holds for the remaining panels.}
\label{fig.emission}
\end{figure*}

The MAGPHYS SED modeling of the sample galaxies performed by \cite{Rubin_2014} provides an
estimate of the dust opacity in the ISM of each system (shown in Table \ref{tab:prop}). 
We make the simplifying assumption
that the wind has the same dust opacity as the ISM, and predict the SBs for an anisotropic wind with this level of dust opacity using the SB reduction factor given above.  These values are shown with blue triangles in Figure \ref{fig.emission}.
For the galaxies J.26, J.36 and J.03, the introduction of dust reduces the predicted emission yet further below our detection limits. For galaxy J.64, in which anisotropy alone did not reduce the predicted emission below our detection limits, Figure \ref{fig.emission} shows that a combination of dust and anisotropy is sufficient to reduce the predicted strength of scattered emission 
so that it is consistent with our observational constraints.

\section{CONCLUSION}\label{sec:conclusion}
We have presented the results of a narrowband imaging search for \ion{Mg}{2} emission around a sample of five star-forming galaxies at a redshift of $z \sim 0.70$ which are known to exhibit outflows traced in \ion{Mg}{2} absorption. We did not detect any \ion{Mg}{2} emission in this sample, and place  upper limits on the surface brightness in the range SB(\ion{Mg}{2}) $< (5.74-6.81) \times 10^{-19}$ ergs sec $^{-1}$ cm$^{-2}$ arcsec$^2$ at 5$\sigma$ significance. These limits are determined within annuli with areas of $\sim 20~\rm arcsec^2$, and having mean radii ranging from 13 to 24 kpc relative to the centers of each target. Our imaging also spatially resolves the strength of the \ion{Mg}{2} absorption observed against the galaxy continua, yielding novel constraints on the \ion{Mg}{2} absorption morphology. 
This absorption fully covers the galaxies from their centers out to isophotal contours defined by the 
1$\sigma$ depth of a continuum + \ion{Mg}{2} image (at 
$R_{\perp}^{\rm SB_1} = 8-21$ kpc),
suggesting that the absorbing gas is optically thick and completely covers the stellar disks out to this distance. Additionally, radial profiles of the mean $\rm EW_{MgII}$ measured for our sample galaxies suggest that the EWs 
are approximately constant across the galaxies' stellar surfaces. 

We compared our surface brightness detection limits with the predictions of the radiative transfer models of \cite{Prochaska_2011}. If the winds in these galaxies do extend beyond the stellar disk, to $\gtrsim 20$ kpc, then we are able to rule out that the winds in our sample are isotropic and dust free, as our images are sufficiently sensitive to detect the emission predicted by such models. Adopting the assumption of dusty and/or anisotropic winds reduces the strength of the predicted \ion{Mg}{2} emission to lie below our detection limits. Although these limits may suggest that the winds in our sample are not isotropic and dust-free, questions linger regarding the relative roles wind anisotropy, dust content, and extent play in reducing scattered emission. Thus, deeper imaging or spatially-resolved spectroscopy of \ion{Mg}{2} will be needed to fully characterize the morphology of these winds. 

\acknowledgements
R.R.V. gratefully thanks Joe Burchett, Karin Sandstrom and Jessica Werk for enlightening discussions which improved this work.
K.H.R.R. acknowledges support from the Alexander von Humboldt foundation in the context of the Humboldt Postdoctoral Fellowship. The Humboldt foundation is funded by the German Federal Ministry for Education and Research.

These findings are in part based on observations collected at the
European Organisation for Astronomical Research in the Southern
Hemisphere under ESO programs 090.A-0427(A).

\newpage
\bibliography{references2017}
\listofchanges
\end{document}